\documentclass[prb,aps,twocolumn,10pt,superscriptaddress,notitlepage,longbibliography]{revtex4-1}
\usepackage{graphicx}
\usepackage{amsmath,mathtools}
\usepackage{amssymb}
\usepackage{epstopdf}
\usepackage{siunitx}
\usepackage{color}
\usepackage{textgreek}
\usepackage[ngerman,english]{babel}
\usepackage[caption=false]{subfig}
\usepackage{bm}

\usepackage{hyperref}
 \hypersetup{
     colorlinks=true,
     linkcolor=blue,
     filecolor=blue,
     citecolor = magenta,      
     urlcolor=red,
     }

\usepackage{braket}
\usepackage{xcolor}

\graphicspath{ {Figures/} }

\usepackage{letltxmacro}
\LetLtxMacro{\ORIGselectlanguage}{\selectlanguage}
\makeatletter
\DeclareRobustCommand{\selectlanguage}[1]{%
  \@ifundefined{alias@\string#1}
    {\ORIGselectlanguage{#1}}
    {\begingroup\edef\x{\endgroup
       \noexpand\ORIGselectlanguage{\@nameuse{alias@#1}}}\x}%
}
\newcommand{\definelanguagealias}[2]{%
  \@namedef{alias@#1}{#2}%
}

\makeatother

\definelanguagealias{en}{english}
\definelanguagealias{EN}{english}
\definelanguagealias{eng}{english}
\definelanguagealias{de}{ngerman}

\begin{document}
	
	\title{Theory of intrinsic propagation losses in topological edge states of planar photonic crystals}

\date{\today}

\author{Erik Sauer}
\email{14ers4@queensu.ca}
\affiliation{Department of Physics, Engineering Physics and Astronomy, Queen's University, Kingston, Ontario, Canada, K7L 3N6}
\author{Juan Pablo Vasco}
\email{juan.vasco@epfl.ch}
\affiliation{Department of Physics, Engineering Physics and Astronomy, Queen's University, Kingston, Ontario, Canada, K7L 3N6}
\affiliation{Institute of Theoretical Physics, \'Ecole Polytechnique F\'ed\'erale de Lausanne EPFL, CH-1015 Lausanne, Switzerland}
\author{Stephen Hughes}
\email{shughes@queensu.ca}
\affiliation{Department of Physics, Engineering Physics and Astronomy, Queen's University, Kingston, Ontario, Canada, K7L 3N6}

\begin{abstract}
    Using a semi-analytic guided-mode expansion technique, we present  theory and analysis of
 intrinsic propagation losses for 
  topological photonic crystal slab waveguide structures with modified honeycomb lattices of circular or triangular holes. 
   Although conventional photonic crystal waveguide structures, such as the W1 waveguide, have been designed to have lossless propagation modes, they are prone to disorder-induced losses and backscattering.  Topological structures have been proposed to help mitigate this effect as their photonic edge states may allow for
   topological protection. However, the intrinsic propagation losses of these structures are not well understood and the concept of the light line can become blurred. 
  For four example topological edge state structures, photonic band diagrams, loss parameters, and electromagnetic fields of the guided modes are computed.
   Two of these structures, based on armchair edge states, are
   found to have significant 
   intrinsic losses
   for modes inside the photonic band gap, 
   more than 100 dB/cm, which is 
  comparable to or larger than typical disorder-induced losses using slow-light
   modes in conventional photonic crystal waveguides, while the other two
   structures, using the valley Hall effect and inversion symmetry, are found to have a good bandwidth 
  for exploiting lossless propagation modes below the light line.
   
\end{abstract}

\maketitle

\vspace{0.3cm}

\section{Introduction}\label{sec:intro}

Semiconductor photonic crystals (PCs) are dielectric structures that allow the manipulation of light on the nanoscale,
achieved by tailoring the periodicity of the dielectric constant \cite{sakoda_optical_2005,patterson_disorder-induced-incoherent_2009,patterson_classical_2009,mannTheoreticalComputationalStudies2017}. In particular, planar photonic crystal slabs (PCSs) have a two-dimensional in-plane periodicity in their lattice structure, which can be used to realize slow light modes on semiconductor chips~\cite{krauss_why_2008}. The PCSs are often introduced with defects within their lattice structures to create waveguides~\cite{notomi_extremely_2001,patterson_classical_2009,vlasov_active_2005,baba_slow_2008,momchil1,notomi2,zhang,mannTheoreticalComputationalStudies2017}, which allow the propagation of light in a particular direction, or trap light in cavities~\cite{akahane_high-_2003,song_ultra-high-_2005,momchil6,noda2,imamoglu,deppe,vuckovic5,notomi3,momchil3,notomi5,noda5,vascoopex}. The fabrication of these PCSs is possible through semiconductor growth techniques~\cite{patterson_disorder-induced-incoherent_2009}, such as etching\cite{kitano_three-dimensional_2015} or lithography~\cite{chan_photonic_2006}.
%
%

In terms of understanding propagation losses,
conventional PCS waveguide structures, such as the 
W1 waveguide (i.e., a single row of missing holes), have been studied extensively \cite{patterson_disorder-induced-incoherent_2009,patterson_disorder-induced-coherent_2009}.
Kuramochi \textit{et al.}~\cite{kuramochi_disorder-induced_2005} have achieved PCS waveguide losses as low as 5 dB/cm,
and O'Faolain \textit{et al.}~\cite{OFaolain:07}
as low as 15 dB/cm. Variations of the W1
design can help improve these
numbers somewhat in terms of reducing the loss
per group index~\cite{Li:08,mann_reducing_2013}.
However, 
in all of these conventional designs, operation
near the mode edge (slow light regime) becomes impractical 
because of significant disorder-induced backscattering~\cite{patterson_disorder-induced-coherent_2009,patterson_disorder-induced-incoherent_2009,mann_role_2015,hughes_extrinsic_2005,Gerace2005,thomas_group_2010,mohamed2,dario3,song,momchil4,vasco_corr}. 

In  recent years, it has been proposed that ``topological''  photonic structures can help mitigate 
the problem of disorder-induced losses in PCS waveguides, thanks to the special properties of their photonic edge states. These edge states of topological waveguides may allow scatter-free propagation for nanoscale PCs and have applications in quantum technologies due to their strong interactions with quantum emitters \cite{anderson_unidirectional_2017,barik_two-dimensionally_2016,parappurath_direct_2018,lu_topological_2014,lu_symmetry-protected_2016,wu_scheme_2015,mehrabad_chiral_2019,paz_tutorial_2020}. 
Experimentally, electromagnetic modes for these topological edge states have been measured by Barik \textit{et al.} \cite{barik_topological_2018} in 2018, indicating that these topological edge states can function as waveguides, which localize spin control. However, for these PCS geometries, the role of out-of-plane losses on the propagating modes is not well understood.
Quantifying such radiative losses is essential to properly characterize the topological edge states in PCSs, and ultimately to improve their
performance and understanding.

To accurately model the behaviour of light within PCSs, numerical solutions
to Maxwell's equations in the full three-dimensional geometry are  required~\cite{patterson_classical_2009,zhao_accurate_2007,andreasen_finite-difference_2008,potravkin_numerical_2012}.
For this purpose, 
well-known numerical approaches, such as the finite-difference time-domain (FDTD) method \cite{taf05} or the plane wave expansion (PWE) method \cite{kim_fourier_2012,Johnson:01}, have been commonly employed during the last two decades. FDTD techniques directly solve Maxwell's equations by iterating through time. Its solutions are numerically exact, however it is a brute-force method which can be computationally inefficient~\cite{patterson_classical_2009,cartar_theory_2017}. This computational inefficiency is especially clear when computing modes above the light line in 3D, and lossy modes can be hard or impossible to resolve with a time-dependent solution. The PWE method, on the other hand, works in the frequency domain rather than in the time domain; PWE solves Maxwell's equations as an eigenvalue problem and is significantly more efficient than FDTD. However, a major limitation with PWE is that it assumes periodicity in all spatial directions, and can only be accurately used for lossless systems and modes, such as standard PCs below the light line~\cite{patterson_disorder-induced-incoherent_2009,joannopoulos_photonic_2008}.

An alternative method to the 
``brute-force'' solvers like FDTD is the semi-analytical method, originally proposed by Andreani and Gerace, known as the guided-mode expansion (GME)~\cite{andreani_photonic-crystal_2006} method. In the GME, the magnetic field of the PCS is expanded in the basis of the guided mode of the slab's effective waveguide, and the resulting eigenvalue equation is solved numerically. The benefits of the GME method are two-fold: (i) it is significantly more computationally efficient than other numerical methods such as FDTD, because the matrix elements of the Maxwell operator become analytical in the guided mode basis; and (ii) the imaginary part of the eigenvalue, which accounts for the the out-of-plane losses, can be obtained by using time-dependent perturbation theory in the low loss regime. This makes the GME an ideal theoretical tool for numerically solving PCSs when the imaginary part of the mode frequency is much smaller than its real part.
Thus, the GME method is an excellent method of choice to analyze the 
photonic band structure and intrinsic propagation losses
of topological PCS  waveguides.

In this work, we study four topological
PCS waveguide designs that have been recently presented in the  literature. 
 The PCS waveguides are analyzed using the GME approach, where we  compute the
 complex photonic band structure and quantify the intrinsic radiation losses above the light line.
We also identify the regions where out-of-plane losses are minimized and characterize the corresponding intensity profiles of the waveguide modes. 
 The first two structures are based on the designs from
Anderson and Subramania~\cite{anderson_unidirectional_2017} and Barik {\em et al.}~\cite{barik_two-dimensionally_2016}, which we show to be intrinsically lossy and form modes inside the
 photonic band gap but above the light line; while the latter two structures,  from 
 Shalaev \textit{et al.}~\cite{shalaev_robust_2019}, and He \textit{et al.}~\cite{he_silicon--insulator_2019}, do have 
 have edge states modes below the light line
 and are thus more promising in terms
 of mitigating problems of intrinsic diffraction losses.

 The layout for the rest of our paper is as follows:
 Section~\ref{sec:designs} introduces
 the main designs of interest, 
 Sec.~\ref{sec:theory} presents the GME theory 
 and methods for computing complex band structure and losses,
 and  Sec.~\ref{sec:results} presents 
 our main results for the four different waveguide designs.
 Our conclusions are presented in Sec.~\ref{sec:conclusions}.
 We also include a Supplementary Material document,
 which contains further details
 of the loss calculations and features, 
 more examples of the various waveguide modes
 found by the GME, and further waveguide mode graphs
to  display the chiral properties of the
 edge state waveguide modes.

 \section{Designs for
 Topological Edge States in Photonic Crystal Waveguides}\label{sec:designs}
 
 \begin{figure}[thb]
    \centering
    \includegraphics[width=0.95\columnwidth]{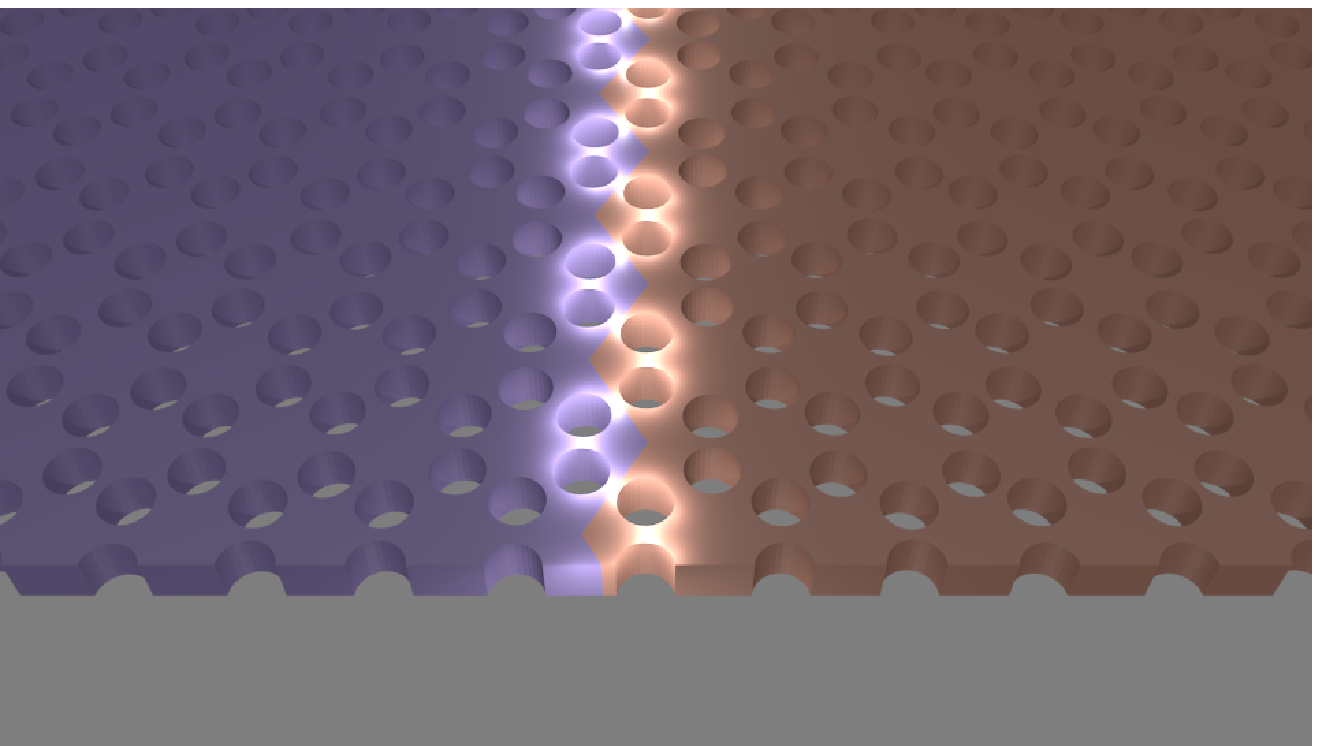}
    \includegraphics[width=0.95\columnwidth]{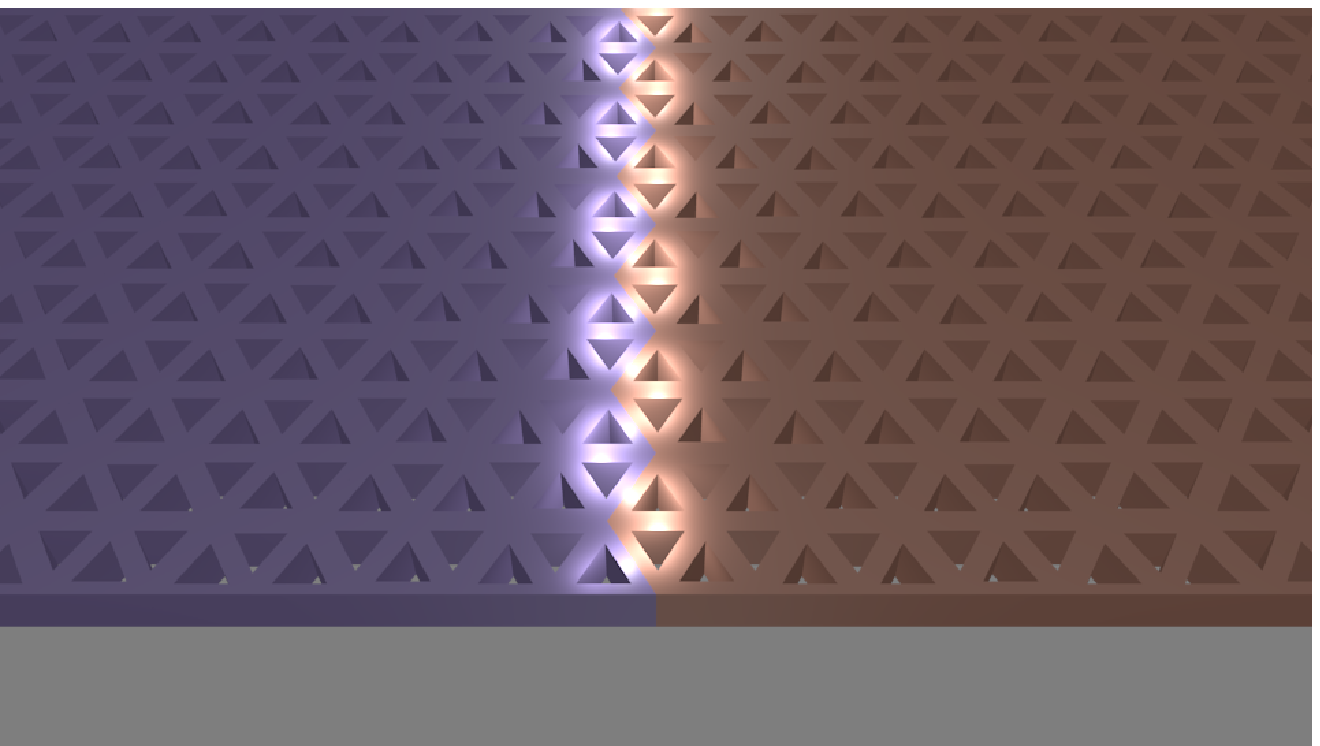}
    \caption{\label{fig:3dModels} Schematic 3D models of two example topological PCS structures: one with circular holes and one with triangular holes. An interface, which acts as a waveguide, separates two lattice structures (in this example, with expanded or shrunken honeycomb lattices).
    }
\end{figure}

 Figure \ref{fig:3dModels} shows two schematic examples  of PCS structures that support
topological edge states, using circles
or triangles on a semiconductor slab (or membrane).
 The interfaces in these two examples separate a topologically trivial lattice structure with shrunken honeycomb clusters and a topologically non-trivial lattice structure with expanded honeycomb clusters.
These two designs, from Refs. \onlinecite{anderson_unidirectional_2017} and \onlinecite{barik_two-dimensionally_2016}, use standard honeycomb lattices, with an armchair interface separating expanded and shrunken honeycomb clusters.
We will refer to these structures as
armchair edge state designs.
Anderson and Subramania~\cite{anderson_unidirectional_2017} 
have presented theoretical photonic band structure calculations for their design of circular holes, as well as power flow diagrams along the interface. However, losses were not
considered.
For the design by Barik \textit{et al.}~\cite{barik_two-dimensionally_2016}, impressive experimental measurements were also demonstrated 
by coupling quantum dot emitters to the waveguide edge states~\cite{barik_topological_2018}. Some partial loss calculations for this structure are available in the Supplementary Material of 
Ref.~\onlinecite{barik_topological_2018}, which are presented in the form of a minimum propagation length
using the 
FDTD technique. 
Experimentally, a loss length of 22 \textmu m was shown for this structure, and they predicted that a loss length of up to 40 \textmu m could be achieved with appropriate parameter adjustments. With such a brute-force FDTD approach, the origin of such losses is not so clear; 
alternative techniques are thus needed not only to highlight the
underlying physics, but also to
explore
parameter space for lower loss designs.

Very recent designs from Refs.~\onlinecite{shalaev_robust_2019} and \onlinecite{he_silicon--insulator_2019} use honeycomb clusters with two alternating hole sizes and instead use inversion symmetry to achieve a waveguide interface. The design proposed by Shalaev \textit{et al.}~\cite{shalaev_robust_2019} is introduced as a topological insulator that  exhibits the valley Hall effect at telecommunications wavelengths. 
A similar design by He \textit{et al.}~\cite{he_silicon--insulator_2019} also exhibits valley Hall effects, and has been experimentally investigated on top of a SiO$_2$ substrate. The design that we will investigate below will not consider the substrate for GME calculations, such that it is consistent with the three other designs, since air bridge structures have less overlap with the light line.
Impressive experiments have been done with both these structures and low losses have been reported.
We will refer to these two structures as
valley Hall edge state designs.

\section{Guided-Mode Expansion Technique and  
Propagation Losses}\label{sec:theory}

For linear and non-magnetic media, one can rewrite Maxwell's equations in the frequency domain, such that a second-order eigenvalue equation in terms of the magnetic field $\bm H(\bm r)$ is obtained:
\begin{equation}
    \label{eq:eigenvalueFull}
    \bm \nabla \times \left [ \frac{1}{\epsilon(\bm r)} \bm \nabla \times \bm H(\bm r) \right ] = \left(\frac{\omega}{c}\right)^2 \bm H(\bm r),
\end{equation}
where $\epsilon(\bm r)$ is the dielectric constant of the slab. To solve this eigenvalue problem using the GME method, the magnetic field is expanded in an orthonormal set of basis states:
\begin{equation}
    \label{eq:expandMag}
    \bm H(\bm r) = \sum_{\mu} c_{\mu} \bm H_{\mu} (\bm r),
\end{equation}
with the orthonormality condition,
\begin{equation}
    \label{eq:magneticOrtho}
    \int_\text{unit cell}  \bm H_\mu^*(\bm r) \cdot \bm H_{\nu} (\bm r) d \bm r = \delta_{\mu,\nu}.
\end{equation}
Then, Eq.~\eqref{eq:eigenvalueFull} is rewritten as a linear eigenvalue problem,
\begin{equation}
    \label{eq:eigen}
    \sum_\nu \mathcal{H}_{\mu \nu}c_\nu = \frac{\omega^2}{c^2} c_\mu,
\end{equation}
where the matrix elements $\mathcal{H}_{\mu \nu}$ are defined as
\begin{equation}
    \label{eq:eigenMatrix}
    \mathcal{H}_{\mu \nu} = \int \cfrac{1}{\epsilon(\bm r)} \left(\nabla \times \bm H^{*}_\mu(\bm r)\right ) \cdot \left (\nabla \times \bm H_\nu(\bm r)\right) d \bm r.
\end{equation}

To solve for $\mathcal{H}_{\mu \nu}$, the GME method obtains the magnetic field for each Bloch wave vector $\bm k$ as a sum of the guided modes over the reciprocal lattice vectors and the mode index $i$. Therefore, the GME for the magnetic field can be rewritten as
\begin{equation}
    \label{eq:gmeMagFull}
    \bm H_{\bm k}(\bm r) = \sum_{\bm G, i} c(\bm k + \bm G,i) \bm H^{\text{guided}}_{\bm k+\bm G,i}(\bm r),
\end{equation}
where ${\bm G}$ is a reciprocal lattice vector for the PCS's lattice structure. The analytical definition for the guided mode $H^{\text{guided}}_{\bm k+\bm G,i}(\bm r)$ varies depending on the slab's layer, and whether the mode is transverse electric (TE) or transverse magnetic (TM) \cite{andreani_photonic-crystal_2006}.
Notice that the matrix elements $\mathcal{H}_{\mu \nu}$ in Eq. (\ref{eq:eigen}) depend on the Fourier transform of the inverse dielectric function in each slab layer $j = \{1,2,3\}$, through
\begin{equation}
    \label{eq:inverseeps}
    \eta_j(\bm G, \bm G') = \cfrac{1}{A} \int_{\text{cell}} \epsilon_j(\bm \rho)^{-1} e^{i(\bm G' - \bm G) \cdot \bm \rho} d \bm \rho,
\end{equation}
where $A$ is the unit cell area, $\bm \rho = (x,y)$, and $j$ represents one of the slab's three layers: the lower cladding, the core and the upper cladding. However, from a numerical perspective, it is much more convenient to calculate the matrix elements of the dielectric function directly as~\cite{andreani_photonic-crystal_2006}
\begin{equation}
    \label{eq:eps}
    \epsilon_j(\bm G, \bm G') = \cfrac{1}{A} \int_{\text{cell}} \epsilon_j(\bm \rho) e^{i(\bm G' - \bm G) \cdot \bm \rho} d \bm \rho,
\end{equation}
and use numerical matrix inversion to find $\eta_j(\bm G, \bm G') = \epsilon_j^{-1}(\bm G, \bm G')$.  This is the approach that we take.

The guided mode basis is computed in an effective homogeneous slab whose dielectric constant is usually taken as the spatial average of $\epsilon_j(\bm \rho)$:
\begin{equation}
    \overline{\epsilon}_j=\frac{1}{A}\int_{\rm cell} \epsilon_j({\bm \rho})d{\bm \rho}.
\end{equation}
Once the magnetic field is obtained from Eq.~\eqref{eq:gmeMagFull}, the electric field is obtained by
\begin{equation}
    \label{eq:electricFromMagnetic}
    \bm E_{\bm k}(\bm r) = \cfrac{ic}{\omega\epsilon(\bm r)} \times \bm H_{\bm k}(\bm r),
\end{equation}
where
\begin{equation}
    \label{eq:electricOrtho}
    \int_\text{unit cell}  \epsilon(\bm r) \bm E_{\bm k}^*(\bm r) 
    \cdot \bm E_{\bm k'} (\bm r) d \bm r
    = \delta_{k,k'}.
\end{equation}

Although performing the GME in this way is 
accurate for photonic modes below the light line, it does not
directly obtain out-of-plane losses.
However, since such losses are small, one can can estimate these losses using perturbation theory. Specifically, 
when a photonic mode escapes the slab's core into the claddings, it couples to lossy radiation modes and falls above the light line. The mode becomes quasiguided and is now subject to intrinsic losses, which can be accurately computed from the imaginary part of the eigenfrequency, $\operatorname{Im}(\omega)$. Similarly to the Fermi's golden rule from quantum mechanics, these losses can be computed by 
second-order time-dependent perturbation theory~\cite{andreani_photonic-crystal_2006}, from
\begin{equation}
    \label{eq:lossesSquared} -\operatorname{Im}\left(\cfrac{\omega^2_k}{c^2}\right) = \pi \sum_{\bm G'} \sum_{\lambda} \sum_{j=1,3} |\mathcal{H}_{\bm k,{\rm rad}}|^2 \rho_j \left(\bm k + \bm G';\cfrac{\omega^2_k}{c^2}\right),
\end{equation}
where $\lambda$ represents either a TE or TM mode, and the matrix element between a guided and lossy radiation mode is given by
\begin{equation}
    \label{eq:radiationmatrixelements}
    \mathcal{H}_{\bm k,{\rm rad}} = \int \cfrac{1}{\epsilon(\bm r)} \left( {\bm \nabla} \times \bm H^*_{\bm k} (\bm r)\right) \cdot \left({\bm \nabla} \times \bm H_{\bm k + \bm G', \lambda, \it j}^{\text{rad}}(\bm r) \right ) d \bm r,
\end{equation}
and $\rho_j$ is the one-dimensional photonic density of states for a given wave vector $\bm g = \bm k + \bm G$ in layer $j$:
\begin{equation}
    \label{eq:DOS}
    \rho_j \left( \bm g; \cfrac{\omega^2}{c^2} \right) = \cfrac{\overline{\epsilon}_j^{-1/2}c}{4\pi}\,\, \cfrac{\theta \left(\omega^2 - \cfrac{c^2 g^2}{\overline{\epsilon}_j} \right)^{-1/2}}{\left(\omega^2 - \cfrac{c^2 g^2}{\overline{\epsilon}_j} \right)^{-1/2}},
\end{equation}
with $\theta$ representing the Heaviside step function. Similarly to those for the guided modes, the analytical definitions for the radiation modes $H^{\text{rad}}_{\bm k+\bm G',\lambda,j}(\bm r)$ depend on the slab's layer and polarization~\cite{andreani_photonic-crystal_2006}.

The imaginary part of the frequency is then easily obtained from ($k$ is implicit)
\begin{equation}
    \label{eq:losses}
    \operatorname{Im}(\omega_k) = \cfrac{\operatorname{Im}(\omega_k^2)}{2\operatorname{Re}(\omega_k)}.
\end{equation}
Subsequently,  the power loss coefficient, $\alpha$, is obtained from
\begin{equation}
    \label{eq:losscoeff}
    \alpha = 2\cfrac{\operatorname{Im}(\omega_k)}{|v_g|},
\end{equation}
where $v_g={d \omega}/{dk}$ is the mode group velocity.
Another useful parameter for connecting to experiments is  power loss in decibels (dB), obtained from 
   $ \text{Loss}_{\text{dB}} = 4.34 \alpha$.

It is worth commenting on the expected accuracy of the GME loss calculations.
As expected from any perturbative approach, the GME method is an approximate one and its predictions are accurate as long as the underlying assumptions are fulfilled. For the photonic dispersion, structures with high-contrast refractive indices are required in order to make a reliable description of PCS eigenmodes with the vertically confined guided mode basis. While for the out-of-plane losses, the imaginary part of the frequency, computed with Eqs.~\eqref{eq:lossesSquared} and \eqref{eq:losses}, must be much smaller than its real part. These two conditions are perfectly satisfied in our cases of interest, where air-bridge high-index slabs are considered and the ratio $\mbox{Im}(\omega)/\mbox{Re}(\omega)$ is on the order of $10^{-4}$ for the largest losses computed below. Nevertheless, structures with weakly confined modes in the vertical direction and with strong radiative losses components are likely not suitable for the GME, and could require first principle simulations, such as FDTD;
such modes are not considered
in this work, and the GME is a far more efficient method to obtain the propagation losses of interest. We also stress that the GME method presents a full vector and 3D solution of 
Maxwell's equations for a wide range of PCS structures.

\section{Numerical Results for the Complex Band Structures and Propagation Losses}\label{sec:results}

In this section, we apply the GME to the four topological PCS structures described in Section \hyperref[sec:designs]{II}, which we will refer to as designs A~\cite{anderson_unidirectional_2017}, B~\cite{barik_two-dimensionally_2016}, C~\cite{shalaev_robust_2019} and D~\cite{he_silicon--insulator_2019}, respectively. For each of the four designs, photonic band diagrams in the $k_x$ direction are computed, along with the nominal light line, and the topological edge states are indicated. A top-down view of each lattice structure's supercell is shown, with the propagation being in the $x$ direction in each case. Since these band diagrams are symmetric about $k_x=0$, only the results for $k_x \geq 0$ are shown.
Zoom-ins of these guided bands of interest are shown, along with their intrinsic losses presented in terms of group index, $n_g = |c/v_g|$, where $v_g$ is the group velocity, and loss coefficient, $\alpha$. Finally, mode profiles of the (normalized) electric displacement field, $\bm D \equiv \epsilon \bm E$, are presented, which provide a visual representation of how well the modes remain confined along the waveguides for their respective topological structure. 

The computational implementation of the GME for the structures below was done in MATLAB. To obtain all necessary results, a choice in the number of $\bm k$ points and the number of basis states were chosen for the dispersion calculations. These numbers are dependent on PCS structure. For design A, a total of 81 basis states and 1002 $\bm k$ points were used. For design B, 144 basis states were computed with a total of 3002 $\bm k$ points. Both design C and D use 60 basis states, with C using 1502 $\bm k$ points and D using 1002 $\bm k$ points. For designs A, B and D, the normalized cut-off in reciprocal lattice vectors, $\bm G a$, was set to 30, whereas it was set to 40 for design C.


\subsection{Armchair Interfaces of Circular Holes by Anderson and Subramania~\cite{anderson_unidirectional_2017} and 
of Triangular Holes by Barik \textit{et al.}~\cite{barik_two-dimensionally_2016}}


\begin{figure}[t]
    \centering
    \includegraphics[width=0.9\columnwidth]{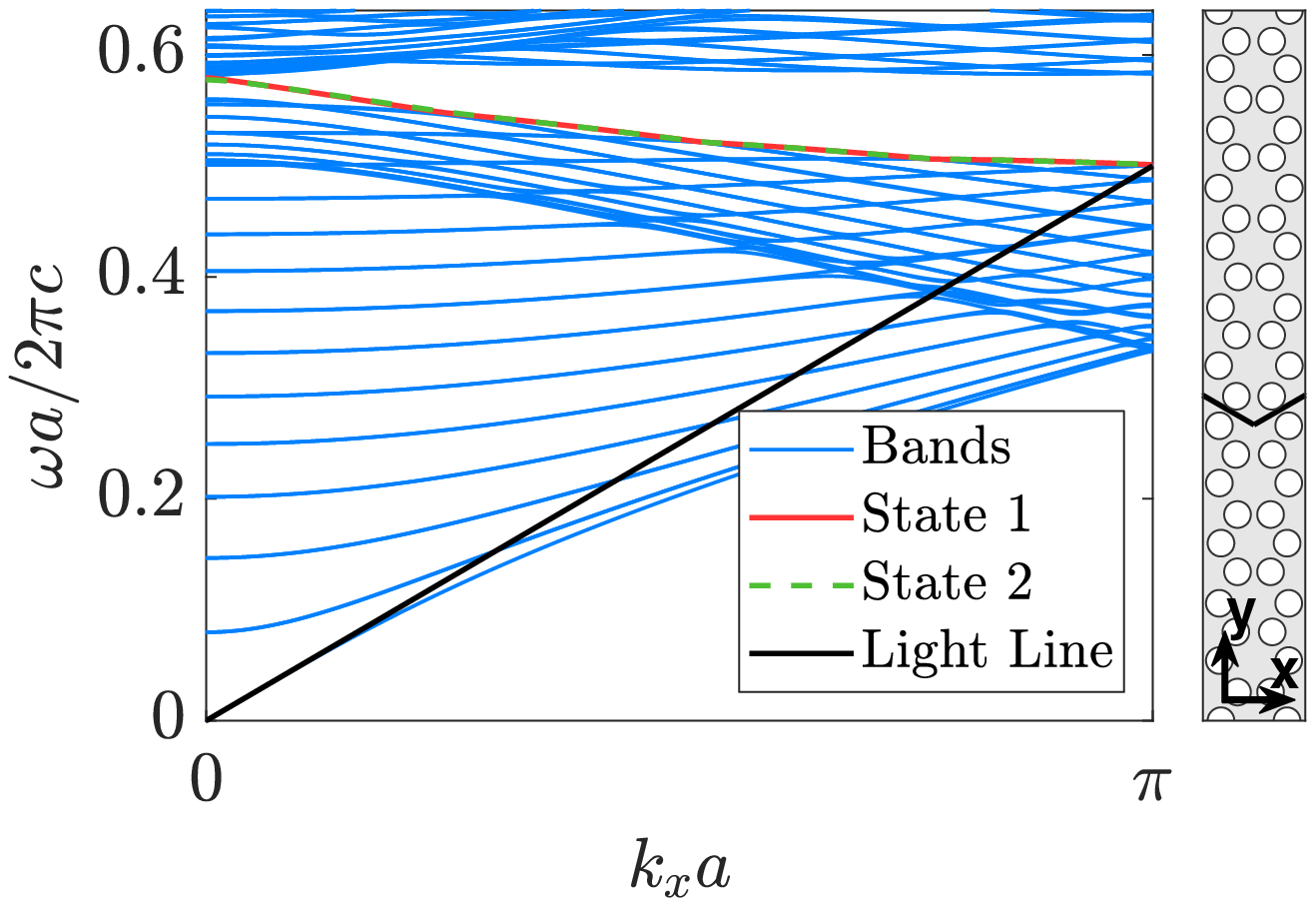}
    \vspace{-0.2cm}
    \caption{\label{fig:disp1}  Dispersion for the lossy armchair interface structure of circular holes, design A~\cite{anderson_unidirectional_2017}. Propagation is along  $x$.}
%
\vspace{0.3cm}
    \includegraphics[width=\columnwidth]{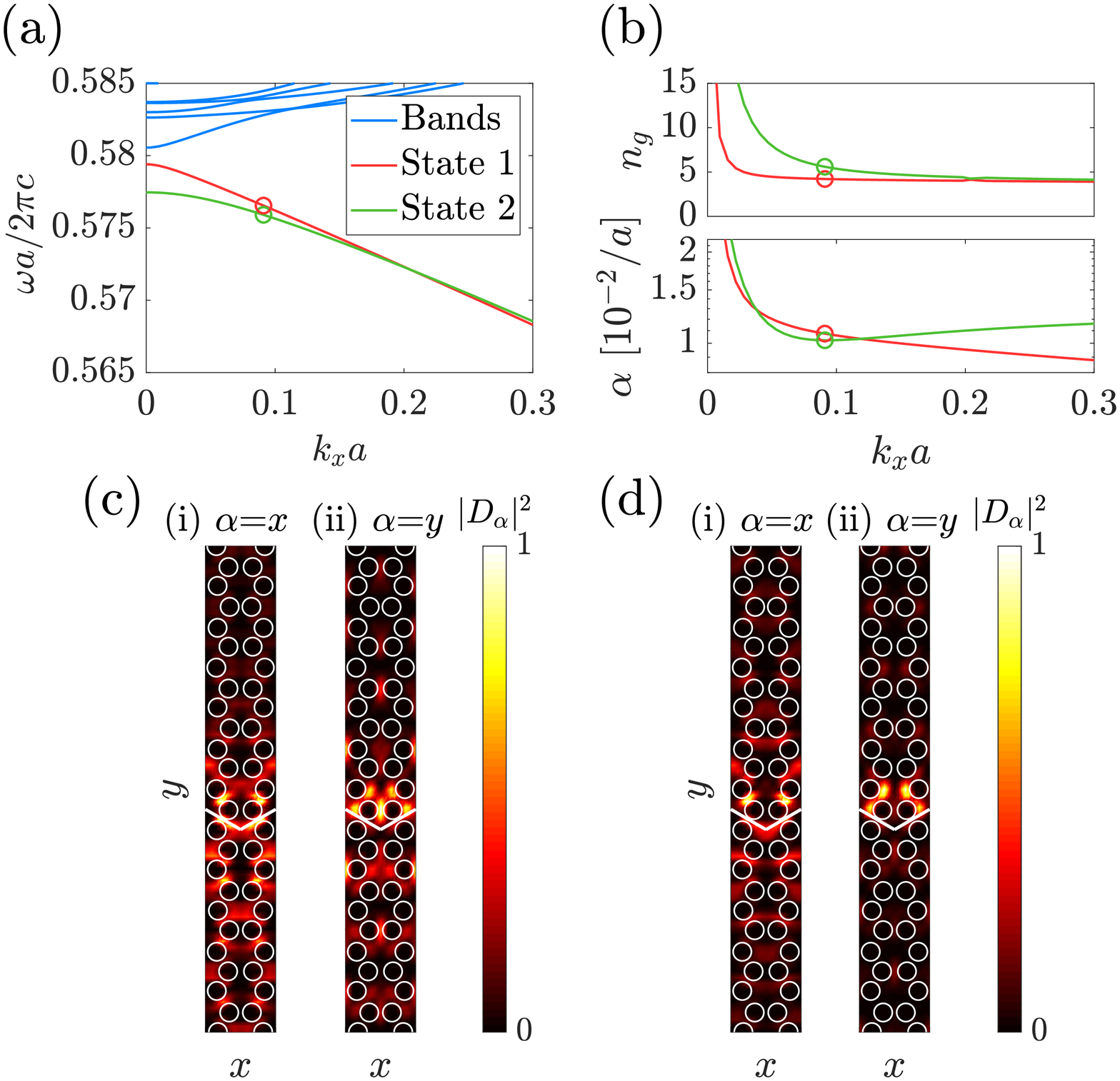}
    \caption{\label{fig:full1_v1}(a) Zoom-in to design A's guided bands of interest, labelled state 1 and state 2. Points of interest, which are found above the light line due to possessing losses, provide the minimum loss coefficient for state 2 (and therefore the maximum loss length) and are indicated by circles. (b) The
    group index (top), $n_g=|c/v_g|$, and loss coefficient (bottom), $\alpha$, of state 1 and state 2, represented with a logarithmic scale. The two points of interest are represented by circles. (c) The electric displacement field mode profiles of the $x$ and $y$ components of state 1 at the point of interest. (d) The electric displacement field mode profiles of
    state 2.}
\end{figure}


\begin{figure}[t]
    \centering
    \includegraphics[width=0.9\columnwidth]{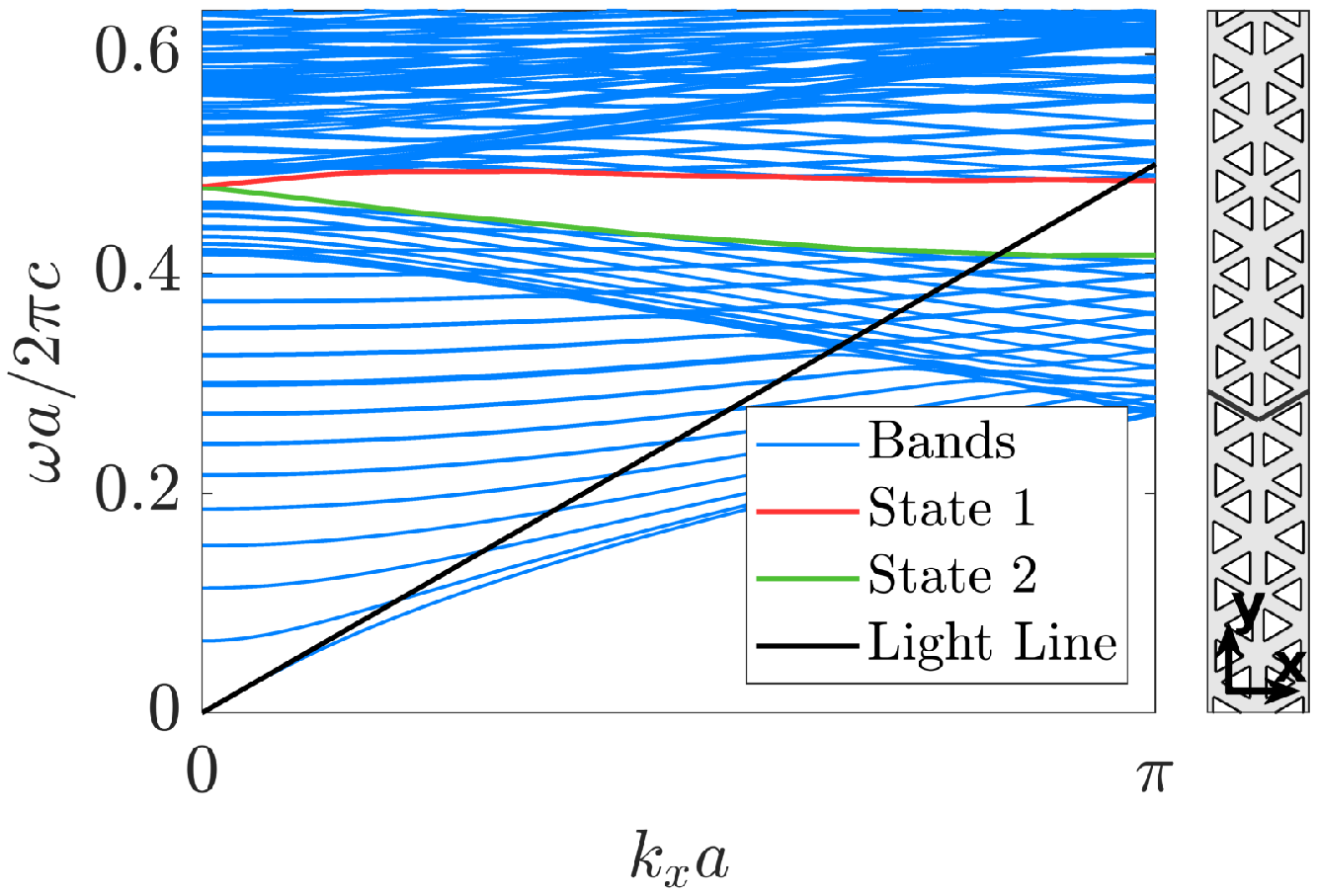}
    \vspace{-0.2cm}
    \caption{\label{fig:disp2}Dispersion for the lossy armchair interface structure of triangular holes, design B~\cite{barik_two-dimensionally_2016}. Propagation is along  $x$.}
%
\vspace{0.3cm}
    \includegraphics[width=\columnwidth]{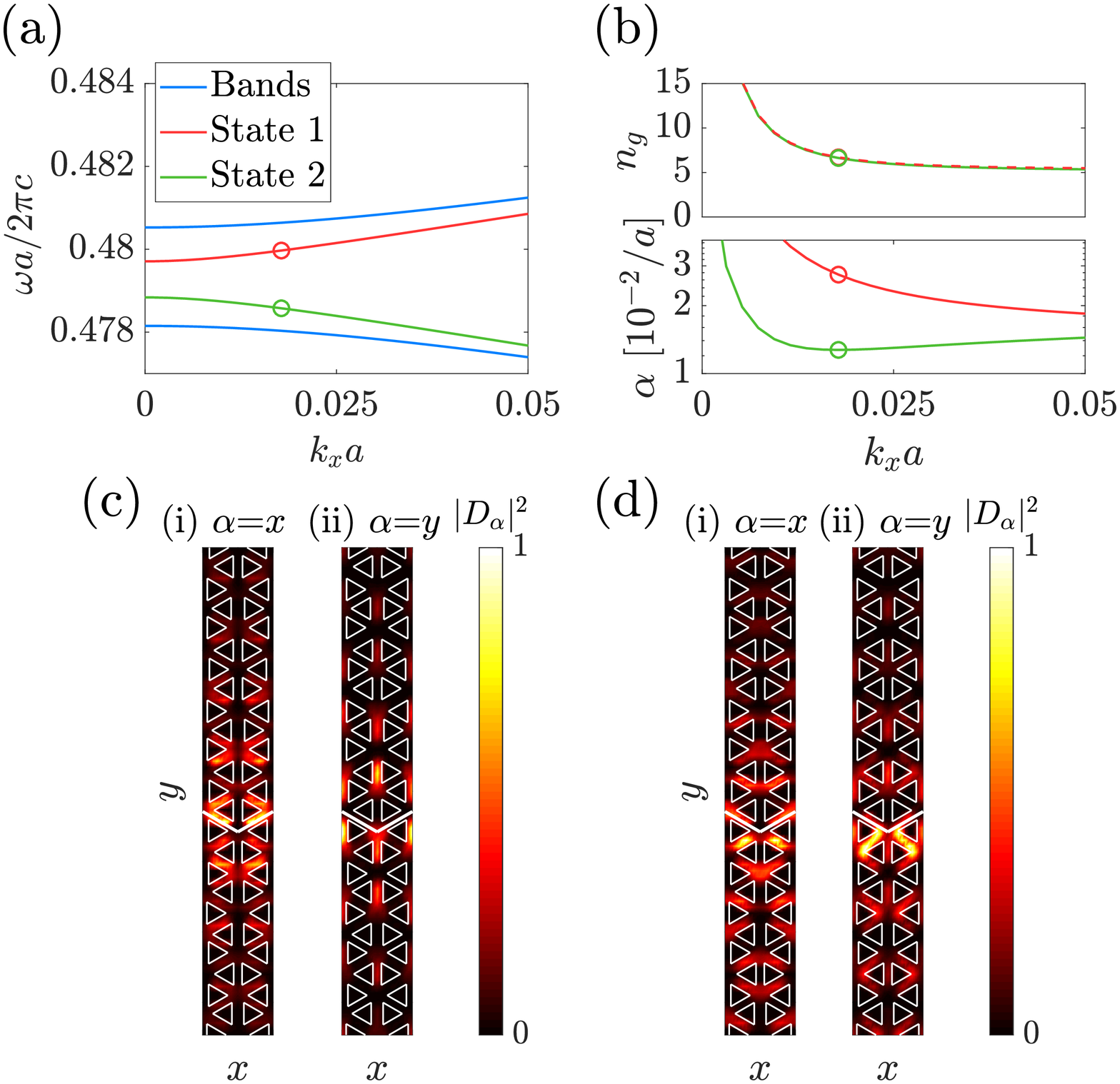}
    \caption{\label{fig:full2}(a) Zoom-in to design B's guided bands of interest, labelled state 1 and state 2. Points of interest, which are found above the light line due to possessing losses, provide the minimum loss coefficient for state 2 (and therefore the maximum loss length) and are indicated by circles. (b) The
    group index (top), $n_g=|c/v_g|$, and loss coefficient (bottom), $\alpha$, of state 1 and state 2, represented with a logarithmic scale. The two points of interest are represented by circles. (c) The electric displacement field mode profiles of the $x$ and $y$ components of state 1 at the point of interest. (d) The electric displacement field mode profiles of
    state 2.}
\end{figure}

We first show results for the PCS structure of the armchair interface of circular holes, introduced by Anderson and Subramania~\cite{anderson_unidirectional_2017}. The GME computations use a slab dielectric constant of $\epsilon_s = 11.5$, a slab thickness of $d = 0.25a$, a hole radius of $r = 0.13a$, and a lattice constant of $a=870$ nm. Assuming the radius of each honeycomb cluster is $R$ and the lattice constant is $a$, the topologically non-trivial side has {\it expanded} honeycomb clusters with $R_{\text{exp}} = a/2.9$ and the topologically trivial side has {\it shrunken} honeycomb clusters with $R_{\text{shr}} = a/3.1$.

The lattice structure and 
 photonic band diagram for this topological structure is shown in Fig.~\ref{fig:disp1}, assuming propagation in the $x$ direction. One might expect the topological edge states in this case to be below the light line, however the GME identifies them to be above the light line, which
 results in  non-zero losses.
It is also important to note that the propagation losses here do not arise from backscattering, but rather from radiation leaking out
vertically while the mode is propagating along the waveguide.

Figure \ref{fig:full1_v1}(a)
shows a zoom-in region of interest of the band structure, highlighting two edge state modes,
labelled as state 1 and state 2. The corresponding
group index and propagation losses
are shown in \ref{fig:full1_v1}(b).
We identify a point of minimum loss
for state 2, at $k_xa = 0.09102$.
The minimum loss coefficient is
$\alpha_{\rm min}=1/97\,a$, which yields
a maximum loss length of $L_{\alpha} = 97 a$.

 Assuming the lattice constant of $a=870$ nm, the minimum losses in this structure were found to be equal to 510 dB/cm. This quantity is
significantly larger 
than typical
disorder-induced losses of conventional
PCS waveguide modes,
which are around 5-30 dB/cm for the fast light
regime, and around 100-1000 dB/cm
for the slow light regime ($n_g \approx 100$)~\cite{kuramochi_disorder-induced_2005,Gerace2005,OFaolain:07,patterson_disorder-induced-incoherent_2009,patterson_disorder-induced-coherent_2009,mann_reducing_2013}. For thin samples, the disorder-induced losses
scale inversely with the group index squared~\cite{hughes_extrinsic_2005,patterson_disorder-induced-incoherent_2009}. The above light-line
intrinsic losses of W1 waveguides
have also been measured to be 
around 400 dB/cm~\cite{kuramochi_disorder-induced_2005}, which is close
to the values of the topological edge states here.


The $x$ and $y$ components of the Bloch-mode displacement fields $\bm D$ 
at $z=0$ (i.e., in the vertical centre of the slab) are shown in Figs. \ref{fig:full1_v1}(c) and (d). These modes, shown for state 1 and state 2 from Fig. \ref{fig:full1_v1}(a), are taken at the points of minimum loss. As expected, the modes remain mostly confined along the interface, however they are still relatively lossy and confinement seems to be somewhat poor for these edge states.

We also studied the results for a similar PCS structure using the zigzag interface of circular holes, also proposed by Anderson and Subramania \cite{anderson_unidirectional_2017}, and find similar results (not shown).
The loss length was found to be equal to $L_\alpha = 166a$, which is equivalent to propagation losses of 173.5 dB/cm, given a waveguide lattice constant of $a\sqrt{3}$. The smaller losses in this structure compared to its armchair counterpart stem mainly from its larger effective mode volume,
$V_{\rm eff}=1/{\rm max}[\epsilon(\bm{r}) |\bm{E}(\bm{r})|^2]$~\cite{manga_rao_single_2007},
which influence the Fermi's golden rule calculations. 
Note this is an effective mode volume per unit cell~\cite{manga_rao_single_2007}.
Although the zigzag structure has somewhat lower losses by approximately a factor of two,
 the modes are still above the light line and
 have significant losses throughout
 all of $k$ space.


We next consider the PCS structure  proposed by Barik \textit{et al.} in 2016 \cite{barik_two-dimensionally_2016},
which was also demonstrated experimentally by coupling spin-charged quantum dots in 2018 \cite{barik_topological_2018}.
The GME computation uses the following parameters: a slab dielectric constant of $\epsilon_s = 12.11$, a slab thickness of $d=160a/445$, a length of one side of the equilateral triangular hole of $L=140a/445$, and a lattice constant of $a=445$ nm. In this case, the topologically non-trivial side has expanded honeycomb clusters with $R_{\text{exp}} = 1.05a/3$ and the topologically trivial side has shrunken honeycomb clusters with $R_{\text{shr}} = 0.94a/3$.

Figure \ref{fig:disp2} shows the photonic band diagram for this topological structure and its lattice design.
Figure \ref{fig:full2}(a)
shows 
a zoom-in of the region of interest, with two 
identified edge state modes.
Once again, we find that these modes are well above the light line
when inside the photonic band gap, though some of the modes
fall below the light line when below the photonic band gap.
Specifically, 
state 1 resides above the light line for $|k_xa| < 3.0495$, whereas state 2 resides above the light line for $|k_xa| < 2.6286$. Within the
photonic band gap region, minimum losses occur at $k_xa = 0.017796$ (for state 2), and the 
corresponding group index and loss
values are shown in \ref{fig:full2}(b).
Here, the loss coefficient achieved at the point of minimum loss is equal to
$\alpha_{\rm min}=1/79\,a$, which corresponds to
a maximum loss length of $L_{\alpha} = 79 a$, and the minimum propagation losses were found to be equal to  $1242$ dB/cm.
Figures
 \ref{fig:full2}(c) and (d) show the components of the guided modes' displacement field for this structure at $z=0$. Similarly to design A, the modes remain mostly along the interface, however the edge state confinement is significantly worse in this case.

It is worth noting that Barik \textit{et al.} have
obtained experimental loss measurements  on this structure and extratced
some values for the optimum loss length \cite{barik_topological_2018,barik_topological_2018}. Using their  lattice constant of $a=445$ nm, our normalized loss length of $L_{\alpha} = 79a$ 
is equivalent to $L_{\alpha} = 35$ \textmu m. Comparing this loss length with their experimental value of $22$ \textmu m,
it is clear that these two values are in reasonable agreement, especially as we have not accounted for any other source of loss, and we have
extracted the theoretical lowest loss as a limit. 
We also expect fabrication imperfections to impact these numbers further.

While both designs A and B produce edge state modes that appear to be intrinsically lossy,
we stress that the physics of these
topological structures is much richer
than regular PC modes~\cite{barik_two-dimensionally_2016,anderson_unidirectional_2017}, and these loss lengths
are certainly large enough to probe many finite-size waveguide effects, exploiting
topology-dependent spin~\cite{barik_topological_2018}.


\subsection{Valley Hall edge state structures by Shalaev \textit{et al.}~\cite{shalaev_robust_2019}
and   He \textit{et al.}~\cite{he_silicon--insulator_2019}}

Next, we examine the  recent PCS edge state
structure of 
 Shalaev \textit{et al.}~\cite{shalaev_robust_2019}, described earlier. The GME computations for this structure use a slab dielectric constant of $\epsilon_s = 12.11$, a slab thickness of $d = 0.639a$, equilateral triangular hole lengths of $L_1=0.4a$ and $L_2=0.6a$, and a lattice constant of $a=423$ nm.
 
\begin{figure}[h!]
    \centering
    \includegraphics[width=0.9\columnwidth]{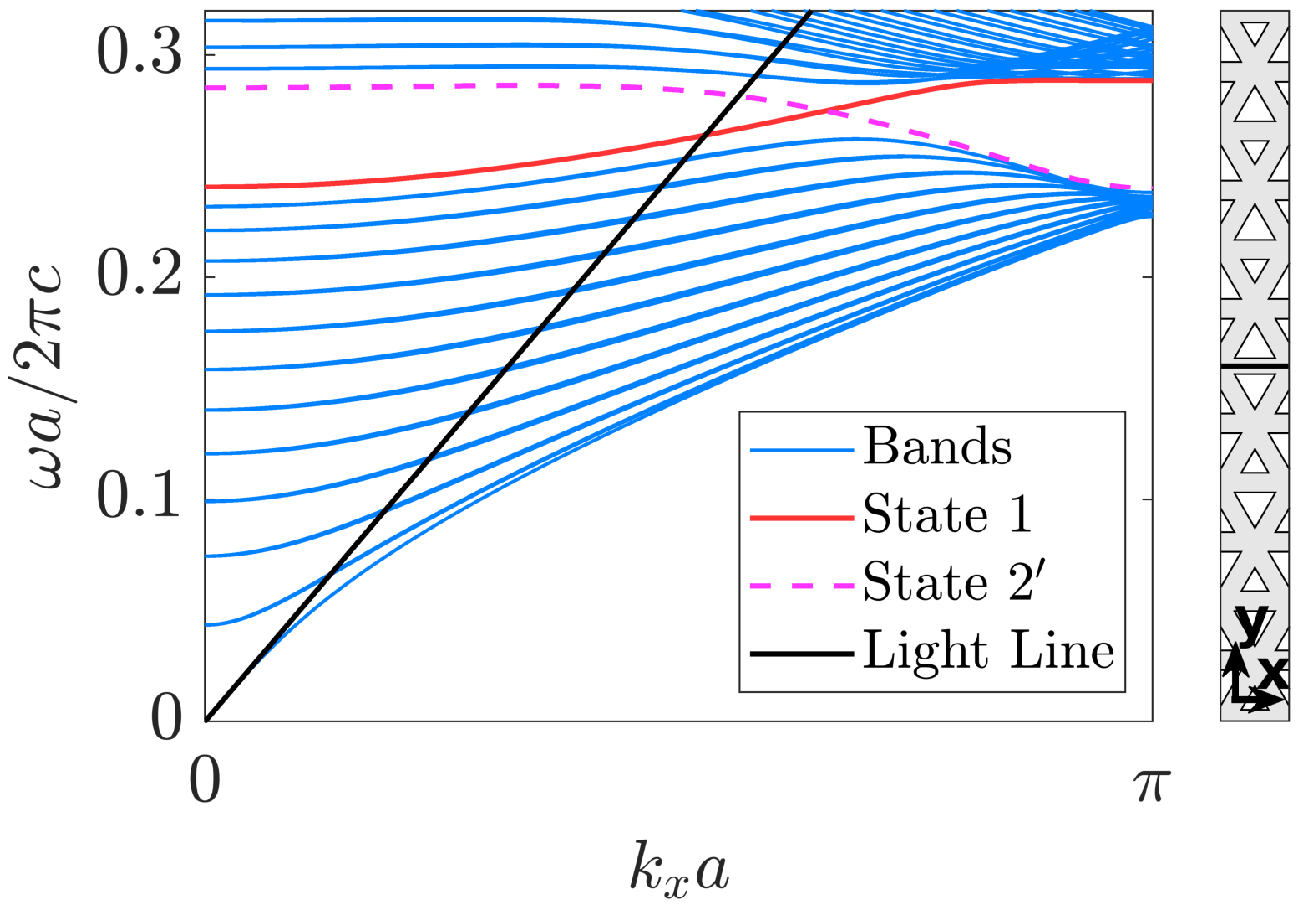}
    \vspace{-0.2cm}
    \caption{\label{fig:disp3}Dispersion for the edge state
structure of triangular holes, design C~\cite{shalaev_robust_2019}. Propagation is along  $x$.}
%
\vspace{0.3cm}
    \includegraphics[width=\columnwidth]{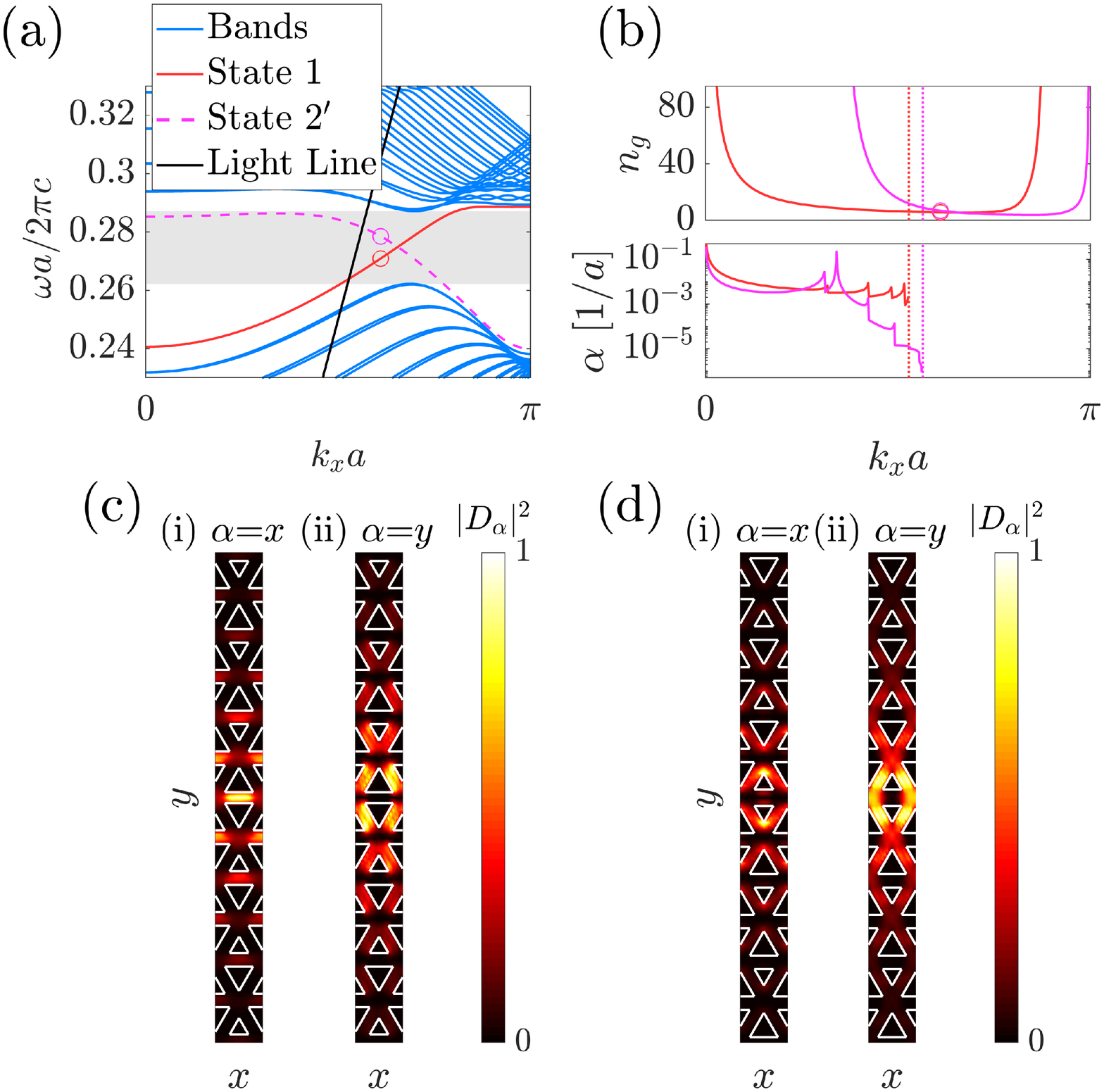}
    \caption{
    \label{fig:full3}
    (a) Zoom-in to design C's guided bands of interest, labelled state 1 and state 2$^\prime$. Note that state 2$^\prime$ is in fact an edge state mode at the border of the supercell, but constitutes an alternative design.
     Points of interest, indicated by circles, were chosen to be below the light line. (b) The group index (top), $n_g=|c/v_g|$, and loss coefficient (bottom), $\alpha$, of state 1 and state 2$^\prime$, represented with a logarithmic scale. The two points of interest are represented by circles, and the light line crossings are indicated by dotted vertical lines. (c) The electric displacement field mode profiles of the $x$ and $y$ components for state 1 at the point of interest. (d) The electric displacement field mode profiles of
    state 2$^\prime$; for simplicity, we show the edge state at the center and rearrange the lattice accordingly (see Supplementary Material).}
\end{figure}

 The full photonic band diagram, as well as the lattice design, for this topological structure are shown in Fig.~\ref{fig:disp3}. Unlike designs A and B, its interface does not separate two honeycomb lattice structures of expanded and shrunken clusters. Instead, a standard honeycomb lattice of triangular holes with two alternating hole sizes has inversion symmetry applied, and an interface is formed from the larger triangular holes. Because of the periodic nature of the GME in both the $x$ and $y$ directions, an intermediate interface of smaller triangular holes forms beyond what is shown in Fig.~\ref{fig:disp3}. The mode confined within the real interface of larger holes is labelled as state 1, whereas the artificial mode confined to the intermediate interface of smaller holes is labelled as state 2$^\prime$; see the Supplementary Material for more information on how these artificial modes (with respect to the original lattice) are defined and obtained.


\begin{figure}[h!]
        \includegraphics[width=0.9\columnwidth]{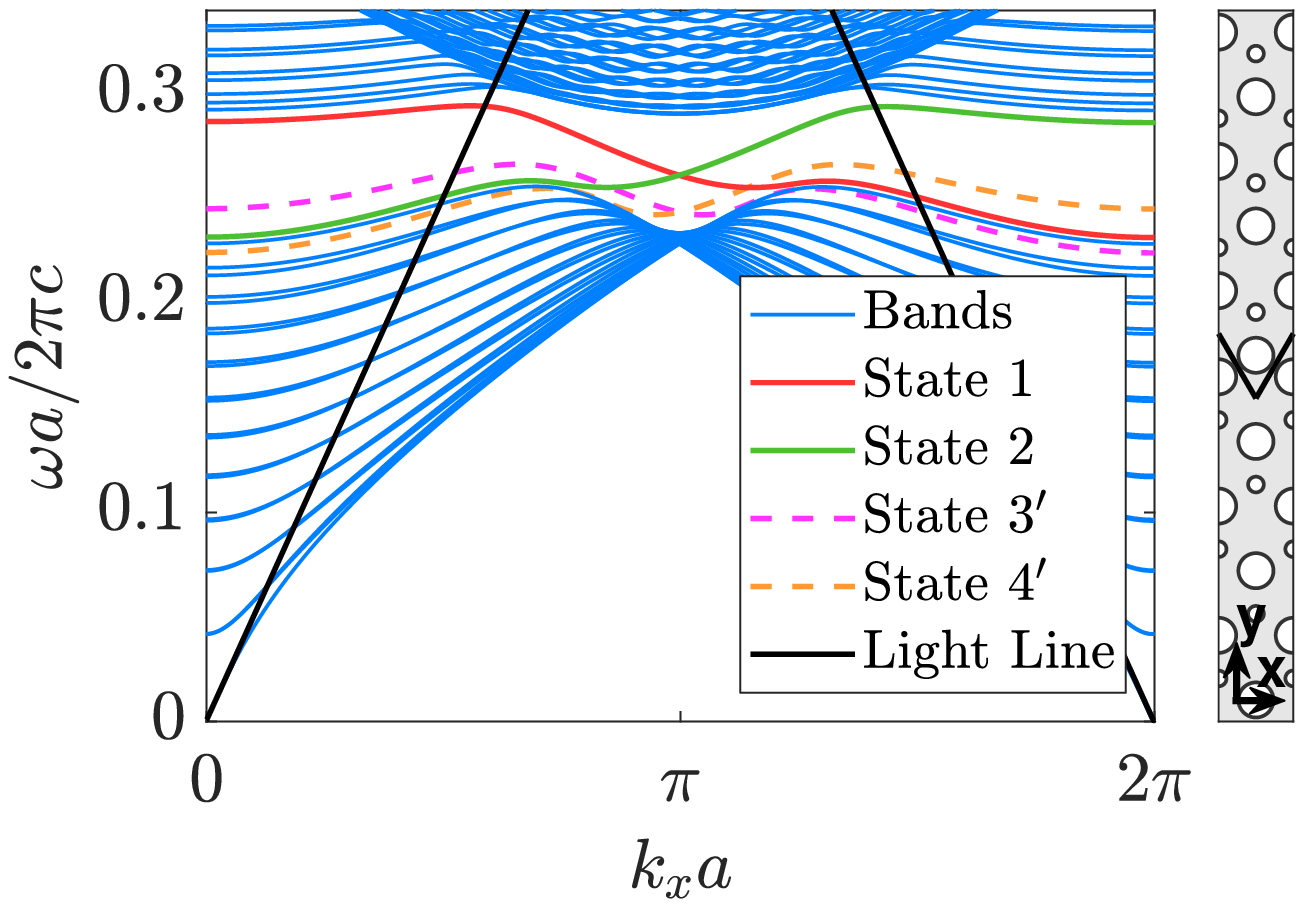}
        \vspace{-0.2cm}
    \caption{\label{fig:disp4}Dispersion for the edge state
structure of circular holes, design D~\cite{he_silicon--insulator_2019}. Propagation is along  $x$.}
%
\vspace{0.3cm}
    \includegraphics[width=\columnwidth]{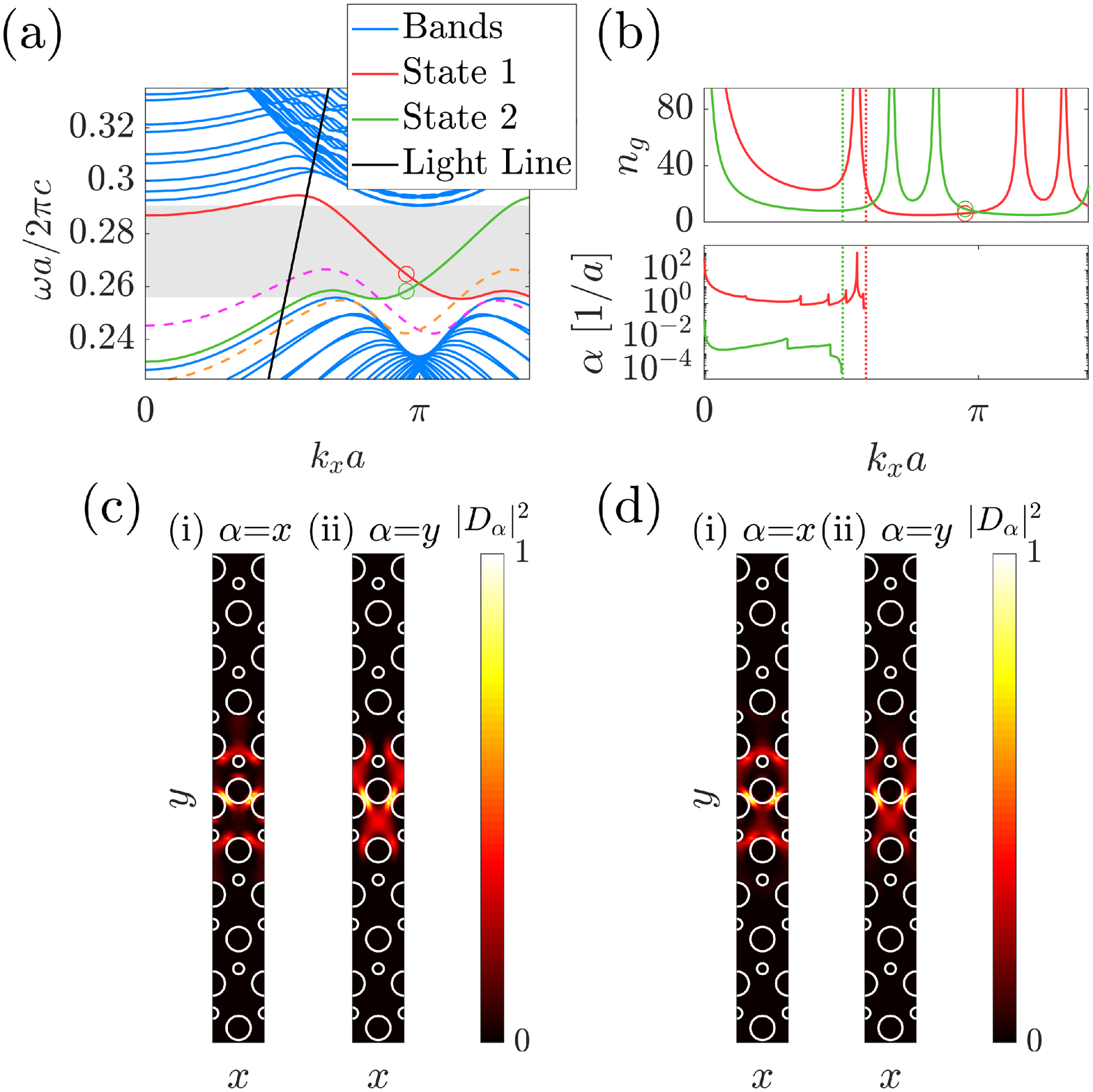}
    \caption{
    \label{fig:full4}(a) Zoom-in to design D's guided bands of interest, labelled state 1 and state 2. States 3$^\prime$ and 4$^\prime$, formed at the border of the supercell, are indicated as dashed bands. Points of interest, indicated by circles, were chosen to be below the light line. (b) The group index (top), $n_g=|c/v_g|$, and loss coefficient (bottom), $\alpha$, of state 1 and state 2, represented with a logarithmic scale. The two points of interest are represented by circles, and the light line crossings are indicated by dotted vertical lines. (c) The electric displacement field mode profiles of the $x$ and $y$ components of state 1 at the point of interest. (d) The electric displacement field mode profiles of
    state 2.}
\end{figure}

Another difference between this structure and designs A-B is that the guided modes lie below the light line, as seen in the zoom-in of the band diagram in Fig.~\ref{fig:full3}(a). Figure \ref{fig:full3}(b) shows the corresponding group index, $n_g=|c/v_g|$, and loss coefficient, $\alpha$ for these two modes. The point at which a mode crosses the light line is indicated by a dotted vertical line. The mode point of interest in this case is arbitrarily chosen to be below the light line. As expected from Fig.~\ref{fig:full3}(b), the loss coefficient, $\alpha$, becomes zero when the mode falls below the light line. The various jumps in $\alpha$
versus wave vector arise from crossings between the Bloch mode and the effective slab waveguide modes (see Supplementary Material). Also, the significant dip in $\alpha$ 
for state 2$^\prime$
is a consequence of the decreasing
$n_g$ and the larger effective mode volume.
As an example, when $\alpha$ of state 1 and state 2$^\prime$ are similar, we obtain effective mode volumes of 0.03183 \textmu m$^3$ (or 0.26375 $(\lambda/n_s)^3$) for state 1, and 0.03752 \textmu m$^3$ (or 0.48910 $(\lambda/n_s)^3$) for state 2$^\prime$; however, when $\alpha$ differs by orders of magnitude, the effective mode volumes increase to 0.03888 \textmu m$^3$ (or 0.39106 $(\lambda/n_s)^3$) for state 1, and 0.05099 \textmu m$^3$ or (0.65045 $(\lambda/n_s)^3$) for state 2$^\prime$, which are comparable to W1-like effective mode volumes \cite{manga_rao_single_2007}.

Figures \ref{fig:full3}(c) and (d) show the components of 
the 
displacement field for both the real and the artificial modes of this structure at $z=0$. Unlike designs A-B, this lattice structure shows a significant improvement in terms of waveguide confinement; the edge state modes remain tightly confined to the interface in this case due to having zero losses.
For the local chiral properties of these modes,
see Supplementary Material.


Finally, we study 
another valley Hall edge state structure
by He \textit{et al.}~\cite{he_silicon--insulator_2019}. For this design, the GME computations use a slab dielectric constant of $\epsilon_s = 12.04$, a slab thickness of $d = 0.571a$, hole radii of $r_1 = 0.105a$ and $r_2=0.235a$, and a lattice constant of $a=385$ nm.
This topological PC design is quite similar to design C; a standard honeycomb lattice of two alternating circular hole sizes has inversion symmetry applied to it, resulting in an interface formed by the larger circular holes. Figure \ref{fig:disp4} displays this lattice structure, as well as the full photonic band diagram. This structure has four significant edge state modes: two real modes labelled state 1 and state 2, and two artificial modes labelled states 3$^\prime$ and 4$^\prime$. Similarly to the artificial mode from design C, states 3$^\prime$ and 4$^\prime$ arise from an intermediate interface formed beyond the supercell length (see Supplementary Material).

Another similarity to design C is the fact that these edge state modes fall below the light line, as seen in the zoom-in of Fig.~\ref{fig:full4}(a). For states 1 and 2,  the group index, $n_g=|c/v_g|$, and loss coefficient, $\alpha$, are shown in Fig.~\ref{fig:full4}(b); the losses for states 3$^\prime$ and 4$^\prime$ are omitted in this case, as the results for states 1 and 2 are more meaningful when comparing to the original waveguide design. Once again, the regions below the light line for this structure provide zero losses, and the jumps in $\alpha$ are due to crossings with the effective slab waveguide (see Supplementary Material). 
For example effective mode volumes,
we obtain
 0.00871 \textmu m$^3$ (or 0.11861 $(\lambda/n_s)^3$) for state 1, and 0.00786 \textmu m$^3$ or (0.09913 $(\lambda/n_s)^3$) for state 2, which are significantly smaller than those of design C. We also show the components of  displacement field for the two main modes of this structure at $z=0$ in Figs.~\ref{fig:full4}(c) and (d).
 Regions of slow light,
 small losses, and small effective mode volumes have applications for on-chip quantum light sources
 including single photon emitters~\cite{manga_rao_single_2007}.
 Similarly to design C, this structure's edge state modes are very tightly confined to the interface. This structure very clearly shows zero losses in these field mode profiles.


\section{Conclusions}\label{sec:conclusions}

In this work, we have applied the guided-mode expansion method to
study four topological photonic crystal slab structures, all of which are modifications to the standard honeycomb lattice structure. Two of these structures, designs A and B,
proposed by Anderson and Subramania \cite{anderson_unidirectional_2017} and Barik \textit{et al.} \cite{barik_two-dimensionally_2016},
consist of  armchair edge states, with an interface separating shrunken and expanded honeycomb clusters. The edge states of these two PCS structures have been shown to fall above the light line, and neither structure seemed to perform particularly well in terms of minimizing propagation loss. Taking previously reported minimum losses of 15 dB/cm and 5 dB/cm for the W1 waveguide as a comparison\cite{OFaolain:07,kuramochi_disorder-induced_2005}, these two topological structures show minimum losses on the order of $10^2$ and $10^3$ dB/cm inside the
photonic band gap. The electromagnetic fields of the guided modes remain mostly along the structures' interfaces, however these edge states are not shown to be tightly confined.

The two other structures that we have analyzed, designs C and D, proposed by Shalaev \textit{et al.}~\cite{shalaev_robust_2019} and He \textit{et al.}~\cite{he_silicon--insulator_2019}, are valley Hall edge state designs that use inversion symmetry to form an interface. The edge states of both PCS structures fall below the light line, thus providing regions of zero intrinsic losses (neglecting imperfections). The electromagnetic field mode profiles confirm that these edge states are tightly confined to their respective structure's interface. Compared to designs A and B, these valley Hall edge state designs seem far superior as a result of their good bandwidth and lossless propagation modes. However, it remains to be quantified how these structures are affected by structural disorder, a topic we will explore in future work.

All of the presented edge state modes show interesting chiral features
for the Bloch mode polarization, which is useful for coupling to spin charged quantum dots
and realizing unidirectional propagation~\cite{Sllner2015,PhysRevLett.115.153901}. Further information on the 
chiral features of the Bloch modes is shown in the Supplementary Material.


\acknowledgements

This work was supported by the Natural Sciences and Engineering Research Council of Canada, Queen's University and the Canadian Foundation for Innovation. 

%

\newpage

\onecolumngrid
\clearpage
\begin{center}
\textbf{\large Supplementary material for ``Theory of intrinsic propagation losses in topological edge states of planar photonic crystals''}
\end{center}
\setcounter{equation}{0}
\setcounter{figure}{0}
\setcounter{table}{0}
\setcounter{page}{1}
\makeatletter
\renewcommand{\theequation}{S\arabic{equation}}
\renewcommand{\thefigure}{S\arabic{figure}}

\renewcommand{\thetable}{S\arabic{table}}



\setcounter{section}{0}
\renewcommand{\thesection}{S-\Roman{section}}


In this supplement,
we provide: (i) further details on the intrinsic loss calculations,
(ii) additional mode plots that
were not shown in the main text,
and (iii) pictures of the chiral features of the 
topological edge state modes studied in the main text.


\vspace{0.3cm}

\section{Further details on the propagation loss features}\label{sec1}


In Figs.~7(b) and 9(b) of the main text, we see several peaks and non-trivial
features in the loss coefficient $\alpha$ as a function of wave vector. 
Apart from the influence of the group index, these effects can be explained by considering the effective slab waveguide, formed by the effective dielectric constant $\epsilon_{\textrm{eff}}$ of the PCS structure. The guided modes of this effective slab are folded when considering the Bloch boundary condition, and the 
photonic crystal slab (PCS) modes will inevitably cross with these guided modes. When these bands cross, the effective slab's guided modes act as a loss channel, thus briefly increasing the losses of the PCS's modes, represented by the frequency's imaginary component $\operatorname{Im}(\omega)$. These peaks are more likely to appear for relatively thicker slabs or those with low contrast in the refractive index.

\begin{figure}[ht]
    \centering
    \subfloat{\includegraphics[width=0.4\columnwidth]{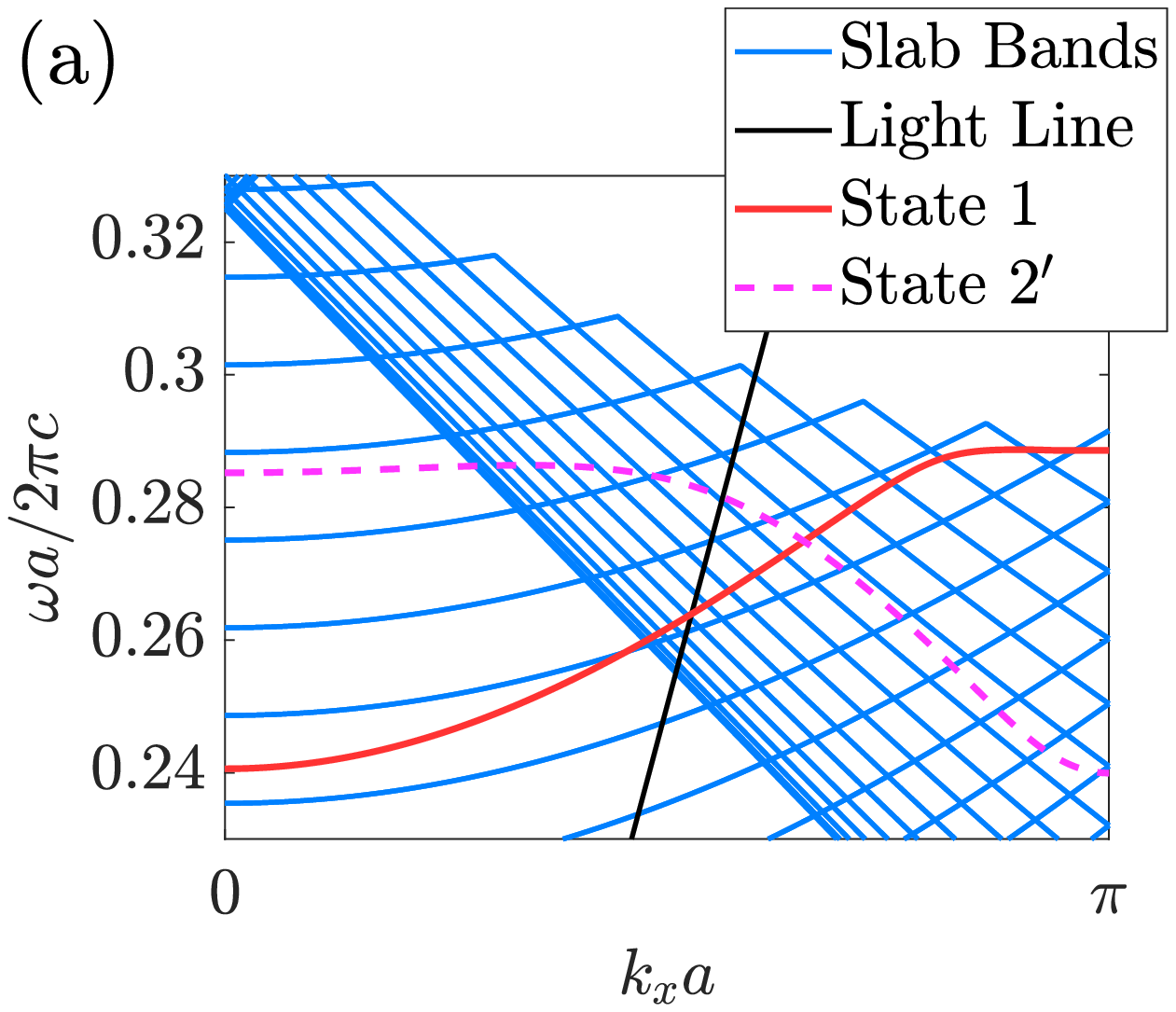}}
    \hspace{1.4cm}
    \subfloat{\includegraphics[width=0.4\columnwidth]{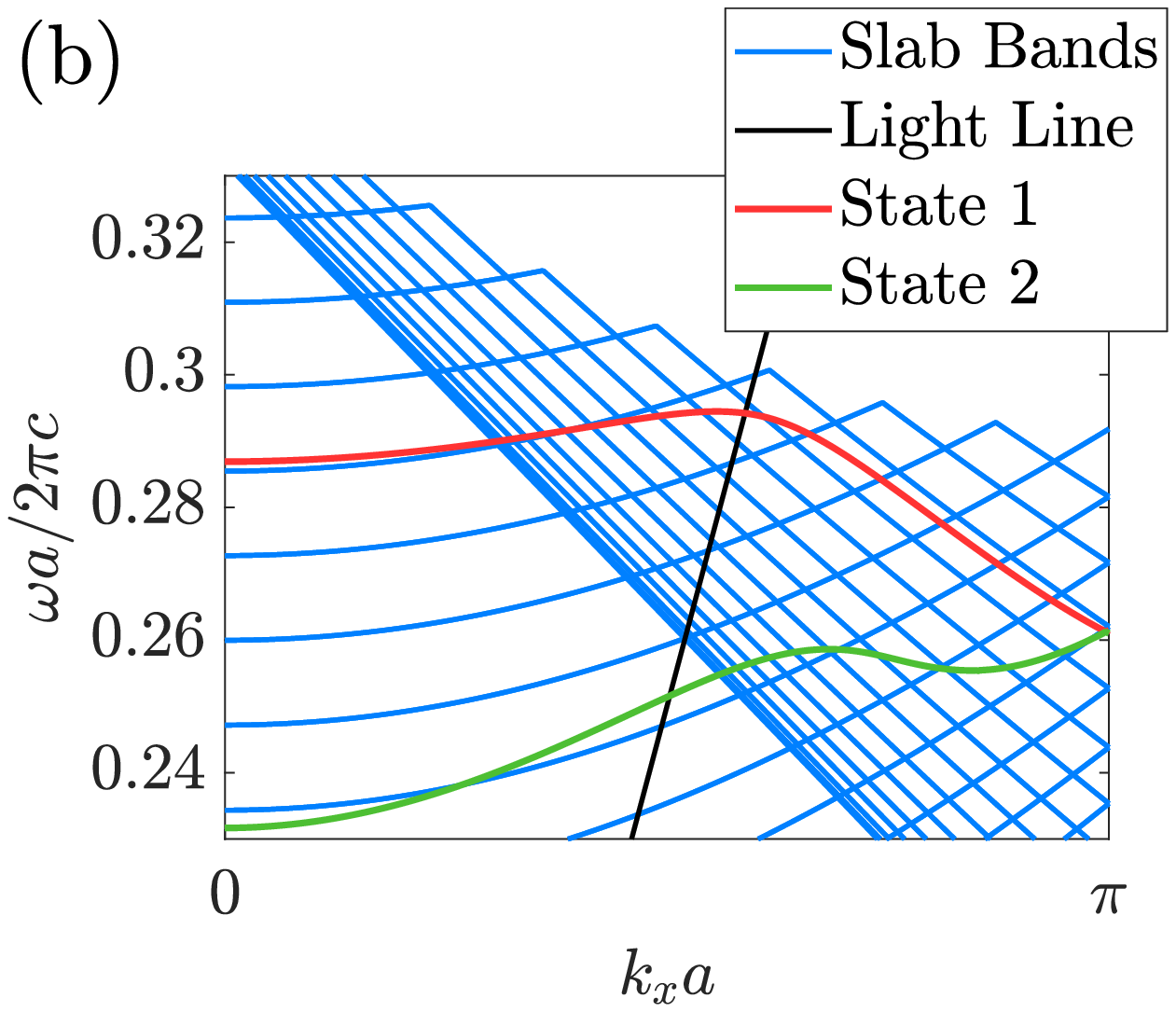}}
    \caption{\label{fig:guidedslabs}Dispersion for the effective slabs of the edge state structures by (a) Shalaev \textit{et al.}~\cite{shalaev_robust_2019} and (b) He \textit{et al.}~\cite{he_silicon--insulator_2019}, designs C and D from the main text, respectively. The two waveguide modes of interest for both structures are overlayed onto this photonic band diagram to show the various instances of band crossings. These crossings result in peaks in $\operatorname{Im}(\omega)$ due to the effective guided slabs acting as loss channels.}
\end{figure}

Figure \ref{fig:guidedslabs} demonstrates this phenomenon for design C and D from Shalaev \textit{et al.}~\cite{shalaev_robust_2019} and (b) He \textit{et al.}~\cite{he_silicon--insulator_2019}, respectively. The two states of interest of these two topological PCS structures are overlayed onto the guided modes of the effective slab to show the locations where the bands intersect, which correspond to the location of the jumps in $\alpha$. Note that the bands shown in this case are only TE-like modes. The other factors that influence the loss calculations
include the group index, $n_g$, and the strength of the normalized Bloch modes
(or the effective mode volume~\cite{manga_rao_single_2007}); larger $n_g$ and smaller effective mode volumes
both increase the loss in a Fermi's golden rule.

\clearpage

\section{Additional Mode Plots and Edge State Modes Identified
at the Edge of the guided mode expansion (GME) Supercell}\label{sec2}

The GME, being periodic in nature, presents specified boundary condition when building two-dimensional lattice structures. Unlike
methods than can apply open boundary conditions,  such as with 
FDTD using perfectly matched layers (PMLs),  the GME builds a lattice structure from an initial unit cell that is repeated periodically in two dimensions. For regular non-waveguide structures, a lattice can be easily built from its clearly defined unit cell. However, waveguide-like structures require a much larger \textit{supercell}, which is in general a rectangular section of the lattice with an interface located at the center. Unlike traditional unit cells, these supercells are only repeated periodically in \textit{one} dimension, which we define as $x$. The $y$ dimension represents the supercell's length, which is assumed to be infinitely long.

Despite these supercells only being periodic in $x$, the GME still views them as being periodic in both $x$ and $y$. Therefore, a supercell length defined as $l_y$ will form interfaces at $y=\pm n l_y$, where $n=\{0, 1, 2, \ldots \}$. 
In principle, we must ensure that $l_y$ is large enough such that the neighbouring interfaces at $y=\pm l_y$ do not interfere with the GME computations. For designs A and B, these periodic interfaces are able to be formed easily, as they are constructed from typical armchair interfaces. Designs C and D, however, utilize inversion symmetry to form an interface, and thus to achieve a perfectly periodic supercell in the $y$ direction, inversion symmetry must be applied \textit{twice}. Doing so results in an intermediate interface located at $\pm (n+1)l_y/2$. The calculations are still physical in the sense that the results have converged, and if we increased the size of the of the supercell, we we could get teh same answer. Thus the GME can find other designs, from the boundary of the supercell.

\begin{figure}[h]
    \centering
    \subfloat{
        \includegraphics[width=0.66\columnwidth]{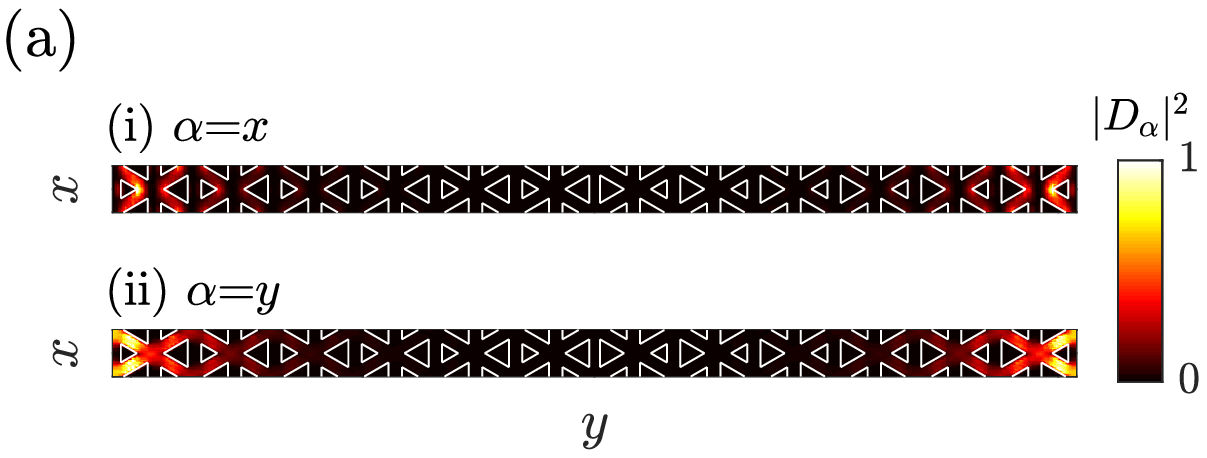}}
    \quad
    \subfloat{
        \includegraphics[width=0.66\columnwidth]{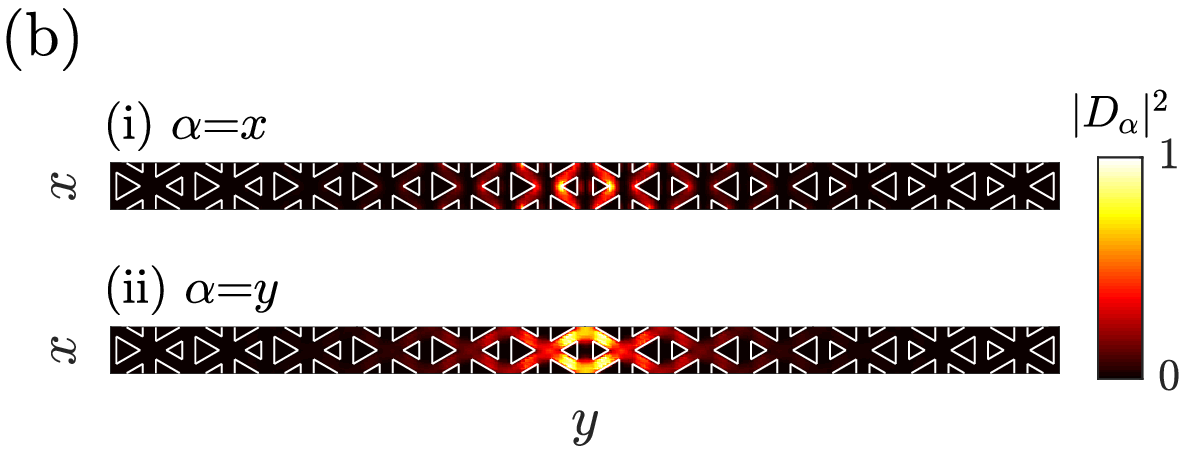}}
    \caption{\label{fig:fields_tri_artificial} Electric displacement field mode profiles for state 2$^\prime$ of the edge state structure by Shalaev \textit{et al.}~\cite{shalaev_robust_2019} (design C in the main text). (a) The full supercell of the real design is shown, demonstrating that state 2$^\prime$ lies at the border of the supercell at an intermediate interface. (b) The lattice structure is altered such that the intermediate interface is at the center of the supercell. State 2$^\prime$ remains confined at this interface nonetheless.}
\end{figure}

Due to the formation of these intermediate interfaces, new modes are formed. These modes, although fictitious from the input design, can be viewed as a real solution to a new design where the intermediate interface is located at $y=0$. One such artificial mode is shown in design C from the main text, labelled as state 2$^\prime$. Figure 2 shows the full supercell of the electric displacement field modes of state $^\prime$ using two different approaches. In Fig.~\ref{fig:fields_tri_artificial}(a), the real lattice design is utilized to demonstrate that the confinement occurs at the edge of the supercell at $\pm l_y/2$, where the intermediate interface is located. An alternative view of this mode is shown in Fig.~\ref{fig:fields_tri_artificial}(b), where the intermediate interface is placed at $y=0$. These two modes are effectively identical, save a translation of $l_y/2$ applied in the $y$ direction.

\begin{figure}[t]
    \centering
    \subfloat{
        \includegraphics[width=0.66\columnwidth]{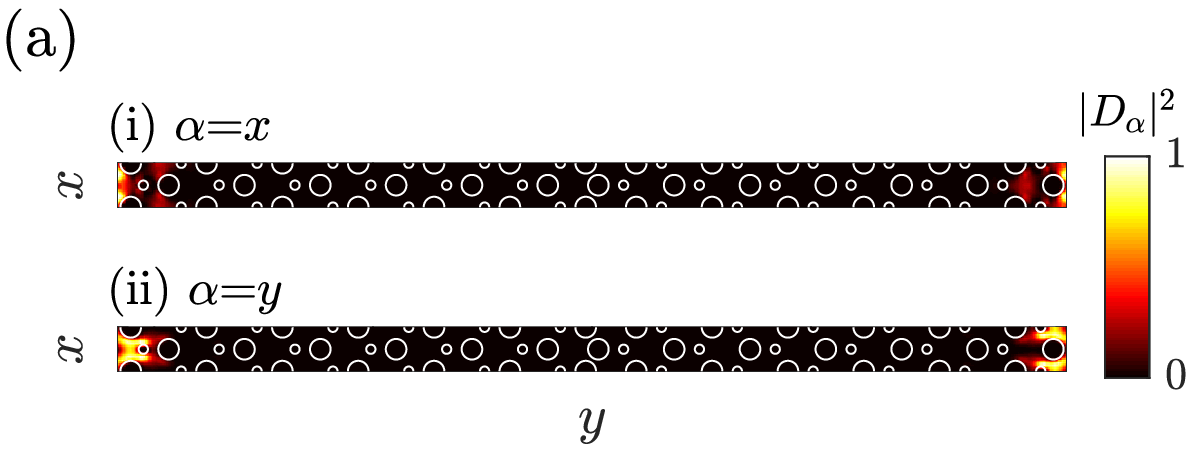}}
        \vspace{-0.5cm}
    \quad
    \subfloat{
        \includegraphics[width=0.66\columnwidth]{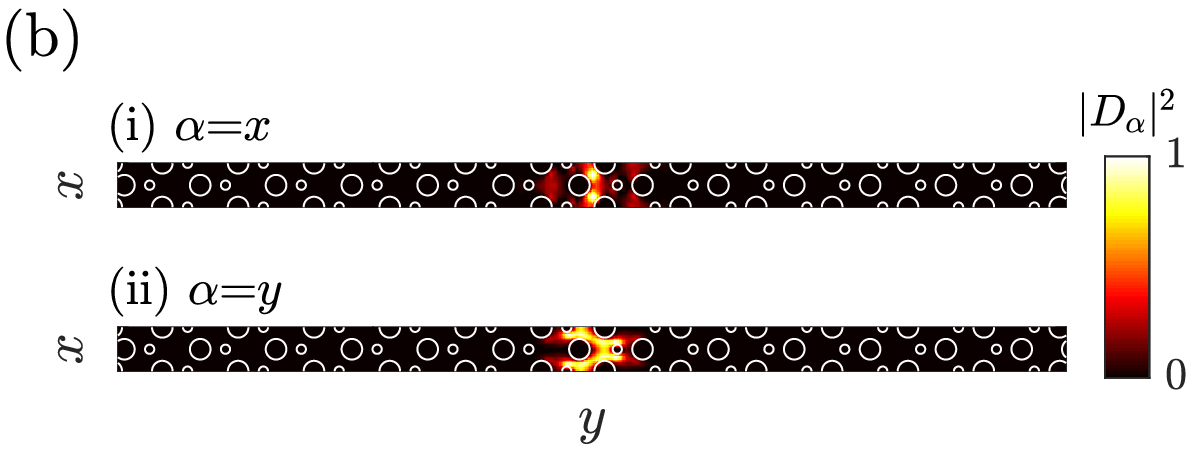}}
           \vspace{-0.5cm}
        \quad
    \subfloat{
        \includegraphics[width=0.66\columnwidth]{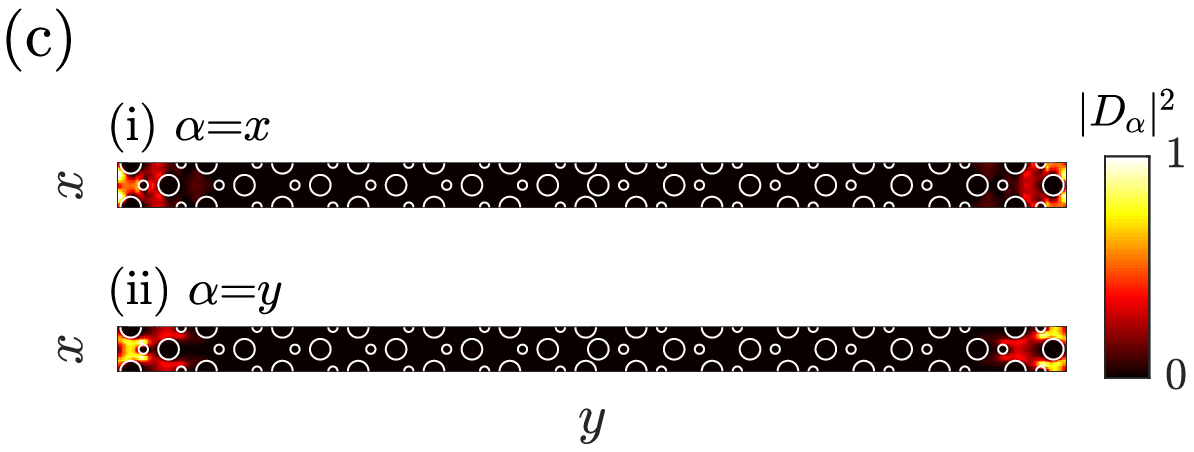}}
           \vspace{-0.5cm}
    \quad
    \subfloat{
        \includegraphics[width=0.66\columnwidth]{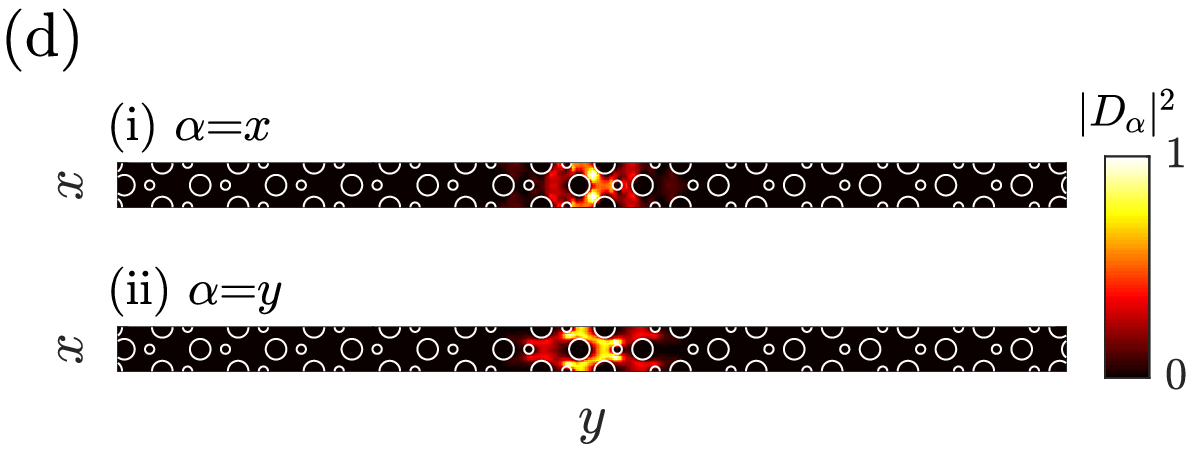}}
        \vspace{0.cm}
    \caption{\label{fig:fields_cir_artificial}Electric displacement field mode profiles for states 3$^\prime$ and 4$^\prime$ of the edge state structure by He \textit{et al.}~\cite{he_silicon--insulator_2019} (design D in the main text). (a) The full supercell of the real design is shown, demonstrating that state 3$^\prime$ lies at the border of the supercell at an intermediate interface. (b) The lattice structure is altered such that the intermediate interface is at the center of the supercell. State 3$^\prime$ remains confined at this interface nonetheless. (c) The full supercell of the real design is shown, demonstrating that state 4$^\prime$ lies at the border of the supercell at an intermediate interface. (d) The lattice structure is altered such that the intermediate interface is at the center of the supercell. State 4$^\prime$ remains confined at this interface nonetheless.}
\end{figure}

The same phenomenon is found in design D of the main text, albeit with two artificial modes rather than one. Figure \ref{fig:fields_cir_artificial}(a-b) shows the electric displacement field mode profiles of state 3$^\prime$ and Fig.~\ref{fig:fields_cir_artificial}(c-d) shows those of state 4$^\prime$. Again, the full supercells are shown in these figures to emphasize the artifical states being located at the edges of the supercells. We show once more alternative views of these modes by wrapping the edges to center, which can be seen as entirely new topological PCS designs.

\clearpage

\section{Chiral Features of the Bloch Modes}\label{sec3}

Here we display the chiral features of the Bloch modes shown in the main text,
to better highlight the regions of circular polarization, which can be used
to couple to spin-charged dipole emitters~\cite{barik_topological_2018,ho_existence_1990} and form one way edge states. Although not necessary for the comprehension of the main text, understanding the chiral properties of these structures has significant value in terms of quantum applications \cite{langStabilityPolarizationSingularities2015,young_polarization_2015}. The method used to show the chiral features of the main text's four structures is through the Stokes' parameters $S_{0,1,2,3}$ as a function of position $\bm r$, which describe the polarization of the electric field $E(\bm r)$:
\begin{equation}
    \label{eq:stokes}
    \begin{aligned}
        & S_0(\bm r) = |E_x(\bm r)|^2 + |E_y(\bm r)|^2, \\
        & S_1(\bm r) = (|E_x(\bm r)|^2 - |E_y(\bm r)|^2)/S_0(\bm r), \\
        & S_2(\bm r) = 2\operatorname{Re}[E^*_x(\bm r)E_y(\bm r)]/S_0(\bm r), \\
        & S_3(\bm r) = 2\operatorname{Im}[E^*_x(\bm r)E_y(\bm r)]/S_0(\bm r).
    \end{aligned}
\end{equation}
The $z$ component of the Bloch mode fields is negligible,
and we will show mode profiles below
as the slab center
only, namely at  $z=0$.

The main benefit of the Stokes' parameters is that they can easily pinpoint locations of polarization singularities, such as circular polarization (``C points'') or linear polarization (``L lines'')
~\cite{young_polarization_2015}.
The first parameter, $S_0$, represents the total electric field strength, which are found from its $x$ and $y$ components. The other three Stokes' parameters,  $-1 \leq S_{1,2,3} \leq 1$, are the three cartesian positions of the Poincar\'e sphere \cite{langStabilityPolarizationSingularities2015}. The parameter of interest here is $S_3$, due to its feature of being able to easily identify C points and L lines. Points where $S_3=\pm 1$ represent C points with left and right circular polarization, respectively; similarly, points where $S_3=0$ depict L lines.

\begin{figure}[ht]
    \begin{minipage}{.45\textwidth}
    \centering
    \subfloat{
        \includegraphics[width=0.42\columnwidth]{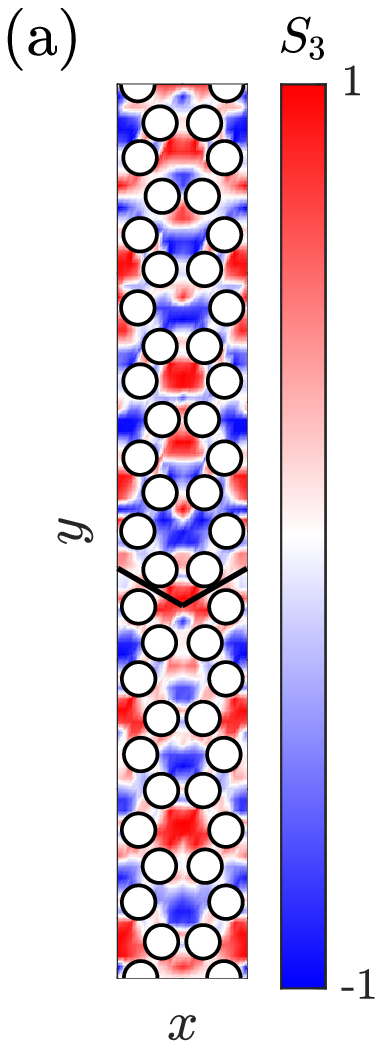}}
    \quad
    \subfloat{
        \includegraphics[width=0.42\columnwidth]{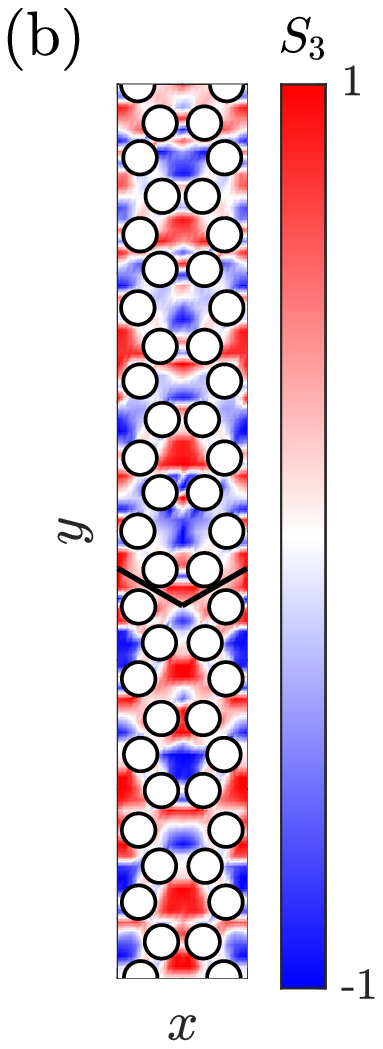}}
    \caption{\label{fig:polarization_A}Polarization for design A's (a) state 1 and (b) state 2, represented by the $S_3$ Stokes' parameter.}
    \end{minipage}
 \hspace*{1cm} \quad
    \begin{minipage}{.45\textwidth}
    \centering
    \subfloat{
        \includegraphics[width=0.42\columnwidth]{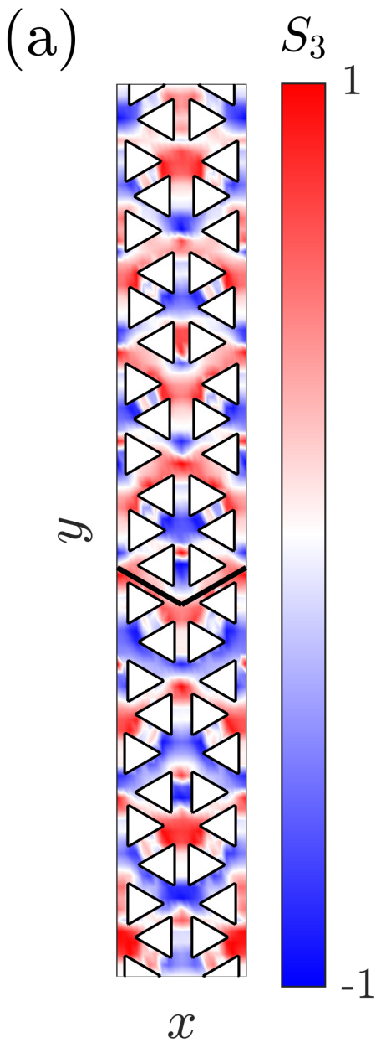}}
    \quad
    \subfloat{
        \includegraphics[width=0.42\columnwidth]{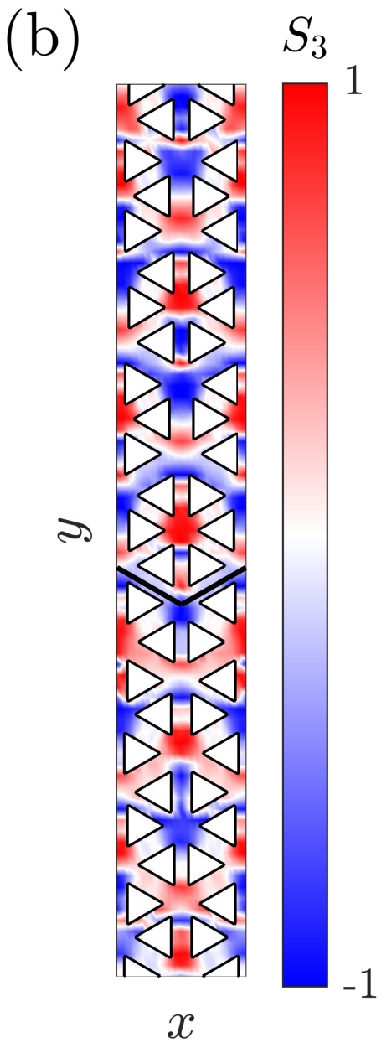}}
    \caption{\label{fig:polarization_B}Polarization for design B's (a) state 1 and (b) state 2, represented by the $S_3$ Stokes' parameter.}
    \end{minipage}
\end{figure}

\begin{figure}[ht]
    \begin{minipage}{.45\textwidth}
    \centering
    \subfloat{
        \includegraphics[width=0.42\columnwidth]{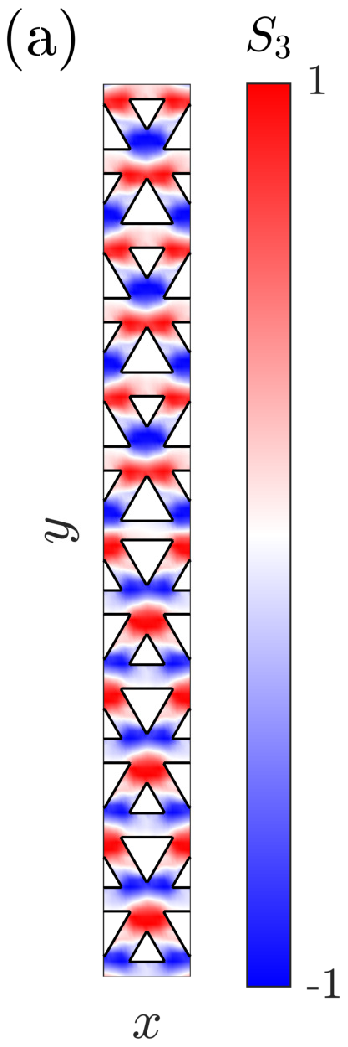}}
    \quad
    \subfloat{
        \includegraphics[width=0.42\columnwidth]{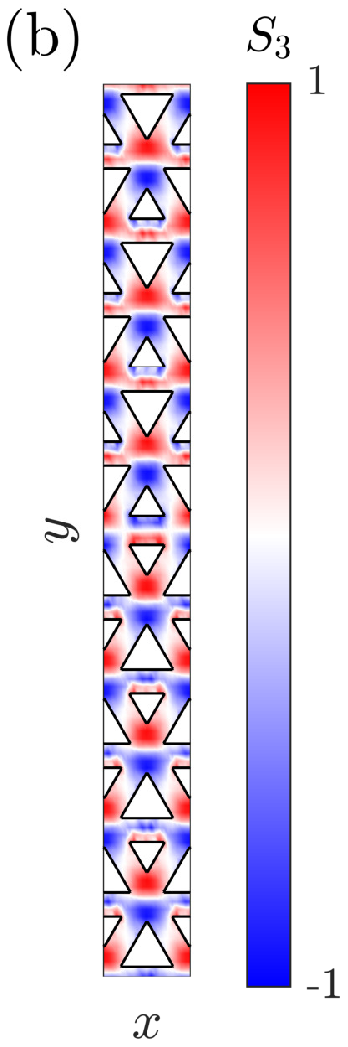}}
    \caption{\label{fig:polarization_C}Polarization for design C's (a) state 1 and (b) state 2$^\prime$, represented by the $S_3$ Stokes' parameter.}
    \end{minipage}
         \hspace*{1cm} \quad
    \begin{minipage}{.45\textwidth}
    \centering
    \subfloat{
        \includegraphics[width=0.42\columnwidth]{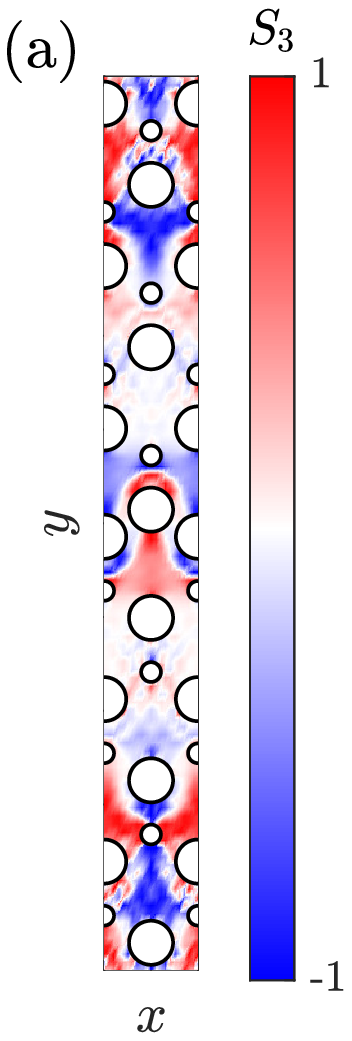}}
    \quad
    \subfloat{
        \includegraphics[width=0.42\columnwidth]{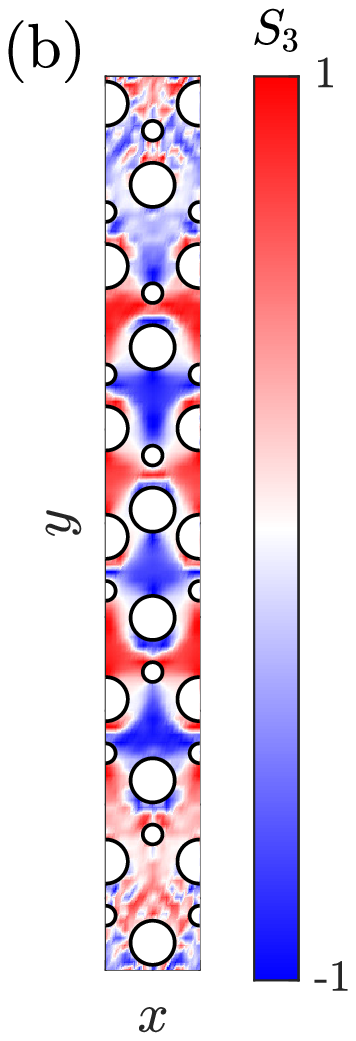}}
    \caption{\label{fig:polarization_D}Polarization for design D's (a) state 1 and (b) state 2, represented by the $S_3$ Stokes' parameter.}
    \end{minipage}
\end{figure}

We show the chiral features of the main text's four structures in terms of $S_3$ to highlight these C points and L lines. Doing so is quite trivial, as the $x$ and $y$ components of the electric displacement field $D$ are available to us. Using the relation $\bm D \equiv \epsilon \bm E$, the Stokes' parameters are easily found for each structure's supercell. The values of $S_3(\bm r)$ for the eight states of interest from the main text are shown in Figs.~\ref{fig:polarization_A}-\ref{fig:polarization_D}, featuring designs A-D, respectively.

The polarizations of the lossy armchair PCS structures (designs A and B) are very much alike due to their similar armchair interfaces. We can see that the lossy armchair PCS structures (designs A and B) tend to preserve both left and right circularly polarized C points within the honeycomb clusters. There do not seem to be any other discernible features for these two structures other than the scarcity in L lines.

For design C, there is a much more even distribution of C points than the previous two structures. However, due to having more locations with L lines, this structure seems to be more versatile. State 1 and state 2$^\prime$ are quite similar, yet state 1's C points are slightly more defined than state 2$^\prime$.

Design D's chiral features are similar to design A and B in the sense that the C points tend to cluster, however they do so around the holes. Additionally, like design C, there are significantly more L lines in this case. It is additionally easier to identify the interface of this structure, as the C-points tend to gather around it. This is consistent with results found by He \textit{et al.}, which demonstrate that left and right circularly polarized C points are found at the center of the interface's holes~\cite{he_silicon--insulator_2019}.

\end{document}